\documentclass[a4paper, fleqn, usenatbib, useAMS]{mnras}

\def\textt{\texttt}

\def\vs{v_{\rm s}}
\def\rt{r_{\rm t}}
\def\rh{r_{\rm h}}
\def\rhh{\rho_{\rm h}}

\def\rp{r_{\rm p}}

\def\rs{r_{\rm s}}
\def\rhs{\rho_{\rm s}}
\def\rhsat{\rho_{\rm sat}}
\def\vs{v_{\rm s}}
\def\vmx{v_{\rm max}}
\def\Tmx{T_{\rm max}}
\def\rmx{r_{\rm max}}
\def\rt{r_{\rm tid}}
\def\rtr{r_{\rm tr}}
\def\ft{F(\mathscr{E})}
\def\nE{\mathscr{E}}
\def\nEt{\mathscr{E}_{\rm tr}}

\def\Mb{M_{\rm b}}
\def\l1{\lambda_1^{\rm p}}

\def\aap{AA}
\def\apjl{ApJL}
\def\apjs{ApJS}

\usepackage{enumitem}
\usepackage{mathrsfs}
\usepackage{afterpage}
\usepackage{multicol}        
\usepackage{graphicx}
\usepackage{float}
\usepackage{bm} 
\usepackage{amssymb}
\usepackage{hyperref}
\usepackage{amsmath} 
\usepackage{subfig} 
\usepackage{pdflscape}
\usepackage{natbib}
\usepackage[export]{adjustbox}
\usepackage{mathrsfs} 
\usepackage{mathtools}

\title[The tidal stripping of CDM subhaloes]{Cold dark matter subhaloes at arbitrarily low masses}
\author[N. C. Amorisco]{Nicola C. Amorisco$^{1}$\thanks{E-mail:
nicola.c.amorisco@durham.ac.uk}\thanks{Ernest Rutherford Fellow}
\\
$^{1}$Institute for Computational Cosmology, Department of Physics, Durham University, South Road, Durham DH1 3LE, UK}

\begin{document}


\pagerange{\pageref{firstpage}--\pageref{lastpage}} \pubyear{2021}

\maketitle

\label{firstpage}

\begin{abstract}
  A defining prediction of the cold dark matter (CDM) cosmological
  model is the existence of a very large population of low mass
  haloes, down to planet-size masses. 
  However, their fate as they are accreted onto 
  haloes many orders of magnitude more massive remains fundamentally 
  uncertain. A number of numerical explorations have found subhaloes 
  to be very resilient to tides,
  but resolution limits make it difficult to explore tidal evolution 
  at arbitrarily low masses. 
  What are the structural properties of heavily stripped subhaloes? Do 
  tidal effects destroy low-mass CDM subhaloes?
  Here we focus on cosmologically motivated subhaloes orbiting
  CDM hosts, and show that subhaloes of any initial mass can be stripped to arbitrarily small mass fractions. 
  We show that previous numerical results can be reproduced by a simple 
  model that describes tidal evolution
  as a progressive `peeling' in energy space and subsequent re-virialization. Tidal heating
  can effectively be neglected 
  at all masses, because its importance i)
   does not increase for subhaloes of 
  decreasing mass at accretion, 
  and ii) it decreases as stripping proceeds. This allows us to predict analytically the structural properties 
  of subhaloes with any initial mass and arbitrary degrees of mass loss. 
  Under the hypotheses that CDM haloes have centrally divergent density profiles 
  and approximately isotropic phase space distributions, our results prove that 
  subhaloes of very low masses are a robust prediction of the CDM model: 
  as for haloes, CDM subhalo populations extend and are abundant down to 
  very low masses. 

\end{abstract}


\section{Introduction}

Cosmological numerical simulations of the process of structure formation show that cold dark matter (CDM) haloes 
are populated by an abundance of gravitationally bound substructures, or subhaloes
\citep[][and references therein]{Springel2005, Frenk2012}. 
Since these substructures are resolved with individual particles, numerical simulations
can only address the existence and properties of subhaloes above a minimum bound mass --
usually taken to correspond to the mass of about 100 particles. It is undisputed 
understanding that CDM simulations find the abundance of bound subhalos to grow with 
a distinctive power law behaviour towards low masses, $dN/dm\propto m^{-1.9}$, all the way down to the mass 
resolution limit \citep[e.g.,][]{Ghigna98, Jenkins2001, Gioc08, Diemand2008, Springel08}. 
Taken at face value, this would suggest very abundant populations of subhaloes 
at even lower masses. But how low?

For a 100 GeV WIMP dark matter (DM) particle, very large populations of isolated 
CDM haloes form with masses as low as the mass of the Earth  
\citep[$\sim10^{-6}$M$_\odot$,][]{Green2005, Wang20}. 
This would suggest that subhalo mass functions may in fact extend down to similarly small masses, and in fact to even lower values as a result of tidal evolution. Needless to say, this is an extremely bold 
extrapolation, and skepticism on its legitimacy is not without motivation. Haloes that 
get accreted by more massive hosts experience strong tidal fields, which strip their mass 
and may cause complete disruption \citep[e.g.,][]{Taylor01,Hay03,Sanders18,Bahe19}.

This fundamental uncertainty on the properties and the very existence of the low-mass 
subhalo populations of CDM haloes undermines a number of research lines that aim 
at using small-scale substructures to constrain the properties of DM itself. Subhaloes 
at the mass scale targeted by gravitational lensing \citep[$\sim 10^{7-10}$M$_\odot$, e.g.,][]{Koo05,Vege12,Hezaveh2016,Gilman2020} 
are resolved or close to being resolved in current high-resolution zoom ins \citep{Richings2021}. 
The subhaloes responsible for disturbing the thin stellar streams of Globular Clusters  
\citep[GCs, $\sim 10^{5-7}$M$_\odot$, e.g.,][]{Ibata02, Johnston02, Erkal2016, Bonaca19} 
are likely 
to be resolved in the near future, despite the difficulties introduced 
by baryonic processes. 
Unfortunately, however, the extrapolation above is unlikely to be bridged fully 
by cosmological simulations in the near future. For a galaxy-sized host with 
$M_{\rm host}=10^{12}$M$_\odot$, a simulation with $N=10^{10}$ particles has a 
mass resolution limit of about $10^4$M$_\odot$, 
10 orders of magnitudes higher than one Earth mass. This 
frustrates those studies that hinge on the existence, abundance and structural properties of subhaloes with even lower masses.
These include the survival of heavily stripped `microgalaxies' in the Milky Way halo 
\citep[e.g.,][]{Err20, Newton2020};
the possibility of DM annihilation, and the associated signal `boost' caused by substructures 
\citep[e.g.,][]{Diem07, Springel08, Del19, Wang20};
possible correlations in the signals from pulsar timing arrays \citep[e.g.,][]{Kashiyama2018, Ramani2020}
the properties and survival of weakly bound wide binaries \citep[][]{Pen19, Pen19b, Kervick21}.

As a consequence, a self-contained and predictive description of the tidal evolution of gravitationally bound collisionless structures would be extremely valuable. 
As highlighted by countless works before \citep[e.g.,][and references therein]{Aguilar85,Taylor01,vdb18}, tidal evolution is a complex non-linear phenomenon, 
in which a number of physical mechanisms operate at the same time. For the case of CDM subhaloes,
at first order, one may ignore i) subhalo-subhalo interactions, which are present 
in the cosmological case but seem to have a secondary role in their evolution \citep{vdb18}; 
ii) the contribution of dynamical friction, which is essentially negligible for subhaloes
of sufficiently low masses \citep[e.g.,][]{White83,Taylor01,Amo17}.
However, there are still tidal heating, tidal stripping, and 
re-virialization, each of which are not easily tackled with analytical methods. This makes a self contained description an elusive objective, especially when looking for a model that can 
be extended to arbitrary initial masses and degrees of mass loss.

Controlled numerical experiments have therefore become increasingly prevalent
in trying to characterize the process of tidal evolution. With regard to CMD subhaloes, 
these studies have highlighted two fundamental results. 

First, that systems with a centrally divergent 
    density cusp, $\rho\sim r^{-1}$, and 
    an isotropic phase space distribution,
    as displayed in CDM cosmological simulations
    \citep{NFW97,Wang20}, are extremely resilient
    to tides \citep[][EN21 in the following]{Kaza04,    Pen08, vdb18, Err20, Err21}. 
    It should be noted that numerical effects -- insufficient force resolution in primis -- 
    cause spurious alterations to the density profile and kinematic properties in the 
    innermost regions of the system, and are therefore responsible for artificial disruption \citep[][]{vdb18,vdb18b}.
    However, when the necessary numerical criteria are satisfied
    \citep{Power03, vdb18b, Og19, Green2019}, subhaloes with the above properties are not 
    observed to fully disrupt above the mass resolution limit of the simulation.
    If the mass resolution allows it and the tidal field is strong enough,   
    subhaloes are seen shedding the largest majority of their mass and still surviving as a self-bound remnant. Complete disruption is only observed as a consequence of limited mass resolution \citep[][EN21]{Kaza04,    Pen08, vdb18, Err20}.
    
Second, that tidally stripped CDM subhaloes populate an essentially uni-dimensional
    sequence, which has been dubbed `tidal track' \citep{Hay03, Pen08}. If the undisturbed progenitor (stripped bound remnant) have the maximum values of their circular velocity 
    profiles $\vmx^0$ ($\vmx$) at the radii $\rmx^0$ ($\rmx$), both dimensionless ratios $\vmx/\vmx^0$ and $\rmx/\rmx^0$ 
    are found to only depend on the bound mass fraction of the remnant itself, $M_{\rm b}/M_0$. It is possible this sequence displays some `thickness', with some numerical 
    results showing it correlates with the subhalos' initial concentration \citep{Green2019}, but the bound mass fraction essentially sets the structural properties of the remnant univocally.

The observed resilience of simulated CDM subhaloes has prompted the suggestion that these are
in fact `indestructible', and can not be fully disrupted by smooth tidal fields \citep[][EN21]{vdb18, Err20}. 
However, without an understanding of the physical reasons underlying this resilience,
it is difficult to justify the extrapolation of  numerical results into possibily different regimes, to arbitrary degrees 
of mass loss and arbitrary satellite-to-host mass ratios. On the other hand, the observation that subhaloes evolve on a simple uni-dimensional tidal track seems to suggest that such an understanding may in fact be attainable. 

It should be stressed that the existence of a  uni-dimensional tidal 
track is an extremely surprising and non trivial result. For instance, and against all 
expectations, the properties of the remnant appear to be independent of its orbit. 
This fact suggests that tidal heating may be a secondary factor in the evolution
of the remnant, as, at fixed host and satellite, the amount of energy injected in the remnant itself is a strong function of its trajectory \citep[e.g.,][]{Spitzer87, Gnedin99}. 
The key reason for this independence is likely connected to the internal dynamical times of CDM subhaloes, a property which has been recently
highlighted by \citet{Err20}. As the dynamical time decreases like $t_{\rm dyn}\propto r^{1/2}$ towards the center of the system, the central parts are effectively 
shielded from tidal shocks, to which they respond adiabatically \citep[e.g.,][]{Spitzer87,Weinberg94,Weinberg94b,Gnedin99}.

Notice however, that the tidal track is also independent of the satellite-to-host mass ratio.
Ultimately, the existence of a tidal track suggests that the properties of the remnant are in fact only dependent on the remnant itself. Remnants with the same bound mass fractions, $\Mb/M_0$, but different initial masses, different hosts or different orbits have identical properties once these are scaled to their initial values. 
This calls for a description of tidal evolution of CDM subhaloes that only involves the subhalo itself.

The simplest approach, however, fails. A simple description of the process of 
tidal stripping of collisionless systems as an effective truncation at the tidal 
radius, $\rt$, does not reproduce the observed phenomenology, independently of the 
specific details. Most evidently, 
it mistakenly predicts tidal disruption for those haloes that are stripped beyond 
the threshold tidal radius of $\rt = 0.77 r_s$ \citep{Hay03}, 
where $r_s$ is the characteristic radius of the initial NFW density profile \citep{NFW97}
\begin{equation}
\rho(r) = {\rhs\over{{r\over\rs}\left(1+{r\over\rs}\right)^2}}.
\label{NFW0}
\end{equation}
Clearly, numerical experiments show otherwise \citep[][EN21]{Kaza04, Pen08, vdb18, Err20}.

An alternative approach is to describe stripping as an effective truncation in 
energy space, rather than in radius. Historically, this is certainly not a new proposal. 
Most notably, 
it motivated the introduction of lowered distribution functions as a model for tidally truncated GCs \citep[e.g.,][]{Woolley54,Michie63,King66,Varri12}. 
More recently, \citet{Choi2009} and \citet{Drakos20} have shown that, 
while a sharp truncation remains an approximation, the stripping of subhaloes 
with centrally divergent density profiles indeed proceeds as an outside-in process 
\citep[see also][]{Read06, Stucker21}. 
Starting from the least bound material, the energy distribution of the system is `peeled' 
in a systematic manner, proceeding towards more bound regions. These results make it attractive to try and describe tidal stripping as a process of progressive energy truncation. This, however, does not complete the model. It remains to be determined 
how to approach the re-virialization that follows the energy peeling. 

Focussing on NFW haloes, \citet{Drakos17} and \citet{Drakos20} have proposed an approach based on the hypothesis that tides cause the remnant to evolve on a sequence of lowered versions of the initial, unperturbed distribution function. Their calculations show tantalising agreement with the results of numerical simulations. However, their
model appears to systematically and significantly overestimate the density in the innermost regions of the remnants, which is worrying when looking for predictions 
that can be extended towards even lower initial masses and bound mass fractions. 
By construction, the model proposed by \citet{Drakos17} addresses re-virialization by 
making explicit choices on both the form of the energy distribution of the system and 
on the quantities that link the latter to the actual density distribution, through the 
Poisson equation. The mentioned mismatch seems to
suggest that some of these choices do not entirely
capture the physics of the tidal evolution 
process.

Here, we follow a perhaps simpler approach. We model tidal stripping as a sharp
truncation in energy space and allow the `peeled' remnants to re-virialize freely
in controlled, isolated N-body simulations. This remains entirely agnostic on the 
dynamics of the re-virialization process. We show that this automatically reproduces
the tidal track. This then allows us to
investigate the regime of arbitrarily small bound mass fractions. In Section 3 we connect
energy truncations to the external tidal field, 
which results in a predictive model of 
tidal evolution for CDM subhaloes with arbitrary initial mass. 
In Section 4 we briefly address the role of the 
central power-law slope of the satellite's density profile. In Section 5 we apply our model to cosmologically motivated CDM halos.
Section 6 sets out the Conclusions.

\section{The tidal track as a sequence in energy truncation} 

In this Section, we focus on satellites with initial density profiles, $\rhsat$, that diverge like $\rhsat\propto r^{-1}$ at the center. In particular, we consider spherically symmetric 
NFW density profiles, as in equation~(\ref{NFW0}), which we exponentially truncate at some virial radius $r_{\rm vir}=c \rs$
\begin{equation}
\rhsat(r) = {\rhs e^{-r/c \rs}\over{{r\over\rs}\left(1+{r\over\rs}\right)^2}},
\label{NFW}
\end{equation}
where $c$ is the satellites' concentration. 
It is useful to also introduce the Dehnen family of density profiles \citep{De93}, which have varying central power-law cusps
\begin{equation}
\rhsat(r) = {\rhs \over{{\left(r\over\rs\right)}^\gamma\left(1+{r\over\rs}\right)^{4-\gamma}}}.
\label{Dehnen}
\end{equation}
The Hernquist model \citep{Hern90} corresponds to the case $\gamma=1$, and therefore shares the same density cusp as the NFW profiles of equation~(\ref{NFW}). Unless otherwise specified, in the rest of this Section we refer to NFW density profiles. Across this study, 
we indicate with $M_0$ the total mass of the unperturbed satellite, 
and with $v_c^0$ its circular velocity profile. 

We start by considering the phase space distribution of the undisturbed
satellite, $f$, which we assume is a function of the mechanical Energy, $E$, alone. This is equivalent to 
assuming that the satellite's phase space distribution is isotropic. In this case, we can recover 
the distribution function $f$ by Abel transform\footnote{We neglect the term in $({d\rhsat/ d\Psi})_{\Psi=0}$ 
as this is zero for all of the density profiles considered in this work.} \citep{Edd16, Widrow00}:
\begin{equation}
f(\mathscr{E}) \equiv {\rhs\over\Phi_0^{3/2}}\ft = {\rhs\over\Phi_0^{3/2}}{1\over\sqrt{8}\pi^2} \int_0^\mathscr{E}{{d^2\rhsat}\over{d\Psi^2}}{d\Psi\over\sqrt{\Psi-\mathscr{E}}} ,
\label{fEdd}
\end{equation}
where $\Phi_0$ is the value at the centre of the gravitational potential of the satellite, 
$\Phi_0=\Phi_{\rm sat}(r=0)$; $\Psi$ is the normalised gravitational potential of the satellite, 
$\Psi=\Phi_{\rm sat}/\Phi_0$; $\mathscr{E}$ is the normalised energy, $\mathscr{E}\equiv E/\Phi_0$.

By definition, the normalised energy $\mathscr{E}$ ranges in the $(0,1)$ interval, 
with $\mathscr{E}\xrightarrow[]{}1$ corresponding to deeply bound material. It is well known 
that, for $0<\gamma<2$, the distribution function $\ft$ diverges as a power-law in the limit $\nE\to1$, with an exponent that is uniquely fixed by the central density cusp \citep[see e.g.,][]{De93, Widrow00}. We have
\begin{equation}
\ft \sim (1-\mathscr{E})^{-{{6-\gamma}\over{2(2-\gamma)}}} .
\label{asymptf}
\end{equation}
The upper panel\footnote{All across this study, we use the symbol $\log$ to refer to the base 10 logarithm.}  of Figure~1 shows the normalised distribution function 
$\ft$ for an NFW density profile with concentration $c\in\{5,10,20\}$.
In agreement with equation~(\ref{asymptf}), the three distribution functions share 
the same asymptotic behaviour as $\mathscr{E}\xrightarrow[]{}1$, and the different 
properties of the truncation at large radii only affect the relative abundance of material towards lower energies.

\begin{figure}
\centering
\includegraphics[width=\columnwidth]{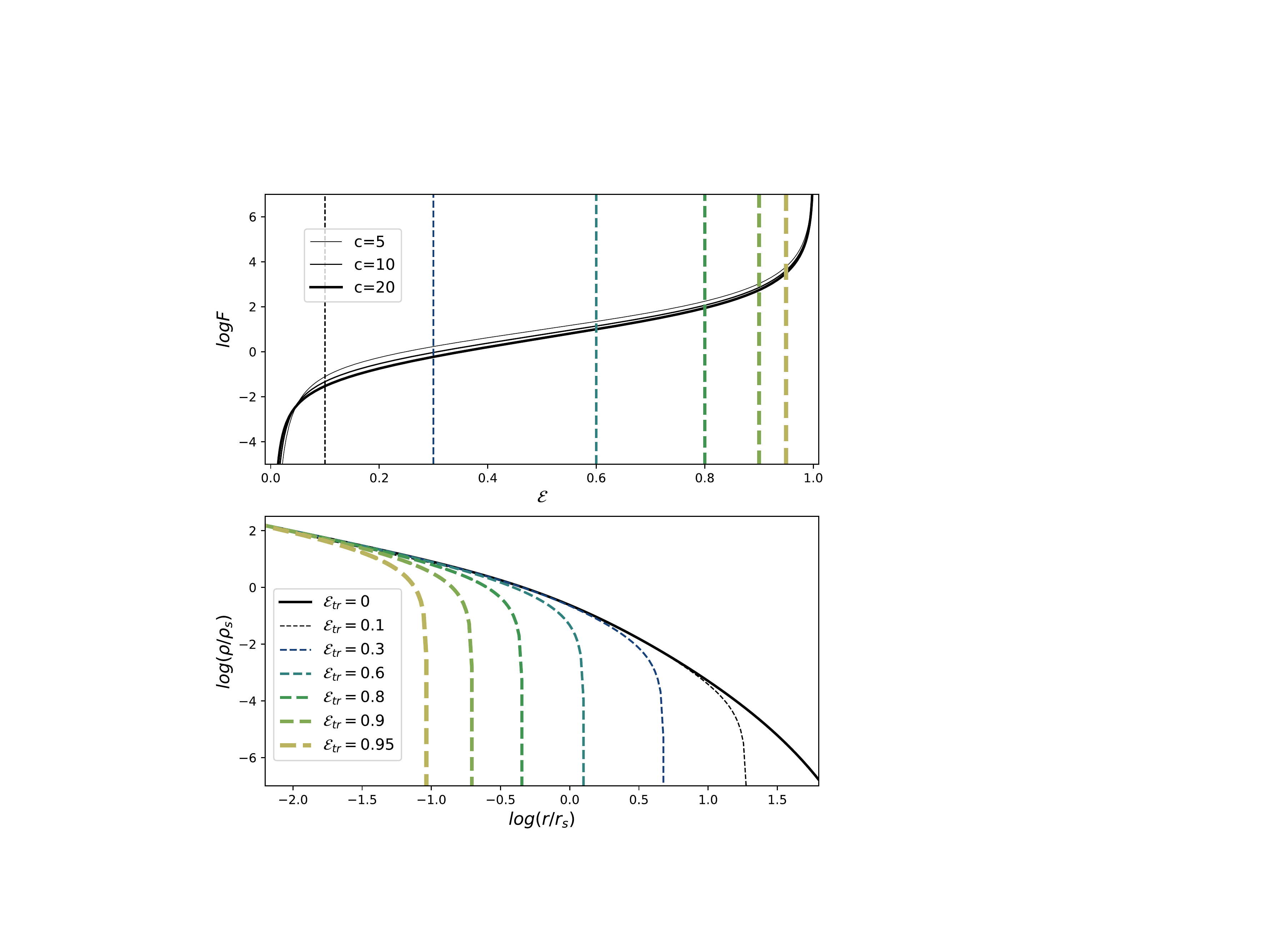}
\caption{Top panel: the dimensionless distribution function for three exponentially
truncated NFW density profiles, equation~(\ref{NFW}), with concentrations $c\in\{5,10,20\}$.
Bottom panel: the density profiles resulting from truncating the system's energy distribution, for the case $c=20$ and different values of the 
normalised truncation energy $\nEt$. 
The same values are displayed as vertical
lines in the top panel.}
\label{fig1}
\end{figure}

\subsection{Energy-trucated halos}

We truncate the distribution function $\ft$ at some truncation energy $\nEt$ by 
simply setting
\begin{equation}
F_{\rm tr}(\mathscr{E}, \nEt) = 
    \begin{cases}
        \ft &{\rm if}\  \nE>\nEt\\
        0 &{\rm if}\  \nE\leq\nEt
    \end{cases} .
\label{ftrunc}
\end{equation}
Such a truncation corresponds to the density profile 
\begin{equation}
\rho_{\rm sat}^{\nEt}(r) = 4\sqrt{2}\pi \rhs \int_{\nEt}^{\Psi(r)} \ft \sqrt{\psi-\nE} d\nE .
\label{rhotrunc}
\end{equation}
In the following, we use the symbol $M^{\nEt}$ to indicate the total 
mass implied by such truncation.

The lower panel of Fig.~1 displays the density profiles associated with different values of the truncation energy $\nEt$. The parent un-truncated density profile is an
NFW profile with concentration $c=20$.
These density profiles are truncated at the radius $\rtr$ where 
\begin{equation}
\Psi(\rtr)\equiv\nEt .
\label{rtrunc}
\end{equation}
The density is identically zero at larger radii
as no material is present at higher energies. Instead, at small radii, the density profiles 
of equation~(\ref{rhotrunc})
are asymptotically identical to the un-truncated original NFW profile: for any truncation energy $\nEt$, it is true that
\begin{equation}
\rho_{\rm sat}^{\nEt}(r)\xrightarrow[r/\rtr \to 0]{} \rho_{\rm sat}(r) = \rhs {\rs\over r} .
\label{rhotruncasy}
\end{equation}
This is the case for any value of the concentration, and is entirely due to the divergence of the distribution function $\ft$ reported in equation~(\ref{asymptf}). While trivial, this asymptotic condition is the first key ingredient of the resilience to tides of satellites with an NFW density profile. 

 Under the hypothesis that tidal stripping proceeds as an outside-in process in energy, the asymptotic identity of equation~(\ref{rhotruncasy}) ensures that the central regions are preserved while the outskirts get stripped.
The stripping process will cause the system to re-virialize, which also affects the central regions. However, no material is physically removed from the innermost regions because of tides.
This is not the case for a cored density profile.
In fact, the central regions of the satellite are
`asymptotically decoupled'. As
all kinematic properties obey asymptotic relations  analogous to 
equation~(\ref{rhotruncasy}), the innermost regions of the system remain in dynamical equilibrium while the outskirts are stripped.
Therefore, the re-virialization process that follows the energy `peeling' also proceeds through equilibrium states of the innermost regions. 

It is interesting to note that these facts are not unique to the $\gamma=1$ case. Equations~(\ref{rhotrunc}) and~(\ref{asymptf}) show that asymptotic
statements like the one of equation~(\ref{rhotruncasy}) remain true as long as $\ft$ diverges as $\nE\to 1$: any density cusp
$\gamma>0$ is sufficient for this. In turn, energy peeling has
fundamentally different consequences for density 
profiles characterized by a finite value of the distribution function $F$ as $\nE\to1$. This is  the case of a cored density profile. If $F$ does not diverge at the center, energy truncation affects the density profile {\it at all radii}. Stripping causes depletion of the central regions,
which find themselves out of both virial and dynamical equilibrium as a consequence of the energy peeling, which in turn facilitates further stripping.

\subsection{The tidal track from re-virialization}\label{numres1}

We simulate the re-virialization of a set of energy-truncated satellites with initial NFW density profiles.
We consider the three cases $c\in\{5,10,20\}$, and for each of them, we focus on the 
set of truncation energies $\nEt\in\{0.36, 0.48, 0.58, 0.68, 0.76, 0.83, 0.89, 0.93, 0.97, 0.99\}$. For the same set of energies, we also examine the case of an Hernquist
density profile. 
We generate initial conditions by sampling the associated distribution functions
with $N=10^5$ particles, independently of the truncation energy. This means that particle masses decrease considerably
for higher values of $\nEt$, as these correspond to smaller values of the total mass $M^\nEt$.
This is shown in the lower panel of Figure~2, which shows the ratio between
the mass of the energy truncated satellites, $M^\nEt$, and the total 
mass of the unperturbed density profile\footnote{The 
masses $M^\nEt$, displayed in Figure~2, include all material in the truncated satellite, both bound and unbound, as defined by equations~(\ref{ftrunc}) and~(\ref{rhotrunc}).}, $M_0$. 

We evolve the truncated satellites in controlled simulations as isolated systems, 
for a total of $10 t_{\rm tr}$, where $t_{\rm tr}$ is the period of a circular orbit 
at the truncation radius in the un-truncated density profile. Simulations are run with 
{\it Gadget2} \citep{Springel05}.
We put special care to making sure that the adopted softening length is such to  
preserve the central density cusp \citep{vdb18b, vdb18}. In our case, this means that we scale the softening length  with the truncation radius of the satellites.

Figure~2 shows the location of the re-virialised truncated satellites in the plane $(\rmx/\rmx^0, \vmx/\vmx^0)$, where $\rmx$ and $\vmx$ are the location and value of the maximum
rotational velocity of the re-virialized satellite, and $\rmx^0$ and $\vmx^0$ are the corresponding values for the unperturbed, un-truncated density profile. 
We measure $\rmx$ and $\vmx$ at the end of our simulations, which provides the 
systems with enough time to settle in new equilibrium states. In fact, the re-virialization 
process is rather quick in the central regions, which stop evolving in an appreciable manner within a handful of dynamical times $t_{\rm tr}$.

Dashed lines in the same Figure display the tidal tracks observed in previous work\footnote{In the interest of reproducibility, we choose not to display the relation proposed by 
\citet{Green2019}. This relation is more complex than the tidal tracks shown in Fig.~2, in that it 
is an explicit function of the instantaneous bound mass fraction of the system. 
Since this is not a quantity that is directly present in our setting at this stage, we can not define a unique track for visualization purposes. The interested reader is referred to 
\citet{Green2019} and EN21, which include direct comparisons between tidal tracks. These are found to be in very good agreement for $\log(\rmx/\rmx^0)\gtrsim-0.6$.}
\citep[][EN21]{Pen08}. Our results appear to delineate a very similar locus. 
It is important to stress the fundamental difference between our results and the tidal 
track. The displayed tidal tracks have been observed by letting NFW profiles evolve
within the tidal field of a host. Our results involve no host or tidal field,
and simulate the re-virialization of an energy-truncated density profile -- 
NFW or Hernquist in this specific case.
Despite this difference, we find that the locus described by our results 
reproduces extremely well the tidal tracks reported in the literature. It is 
worth noticing that published tidal tracks differ from each other for
$\log(\rmx/\rmx^0)\lesssim-0.6$. Smaller values of the ratio $\rmx/\rmx^0$ correspond
to bound mass fractions $\Mb/M_0\lesssim 10^{-2}$. This strongly suggests that these mismatches
originate from numerical effects, which, in turn, may operate differently in different numerical
implementations (see also EN21). 

Since our approach implies that the different satellites are 
resolved with the same number of particles, our results extend to significantly
smaller values of the ratio $\rmx/\rmx^0$, corresponding to higher degrees of mass loss. 
In particular, we find that our locus scales like $\vmx/\vmx^0\sim (\rmx/\rmx^0)^{1/2}$
for small values of $\rmx/\rmx^0$, a result on which we return in the Section~\ref{modelasy}.
The gray-dashed line in Figure~2 is a parametric curve which closely follows
previously proposed tidal tracks in the regime of `low' mass loss, $\log(\rmx/\rmx^0)\lesssim-0.6$, and reproduces our results in the regime of high mass loss. A description is available in Appendix A.

\begin{figure}
\centering
\includegraphics[width=\columnwidth]{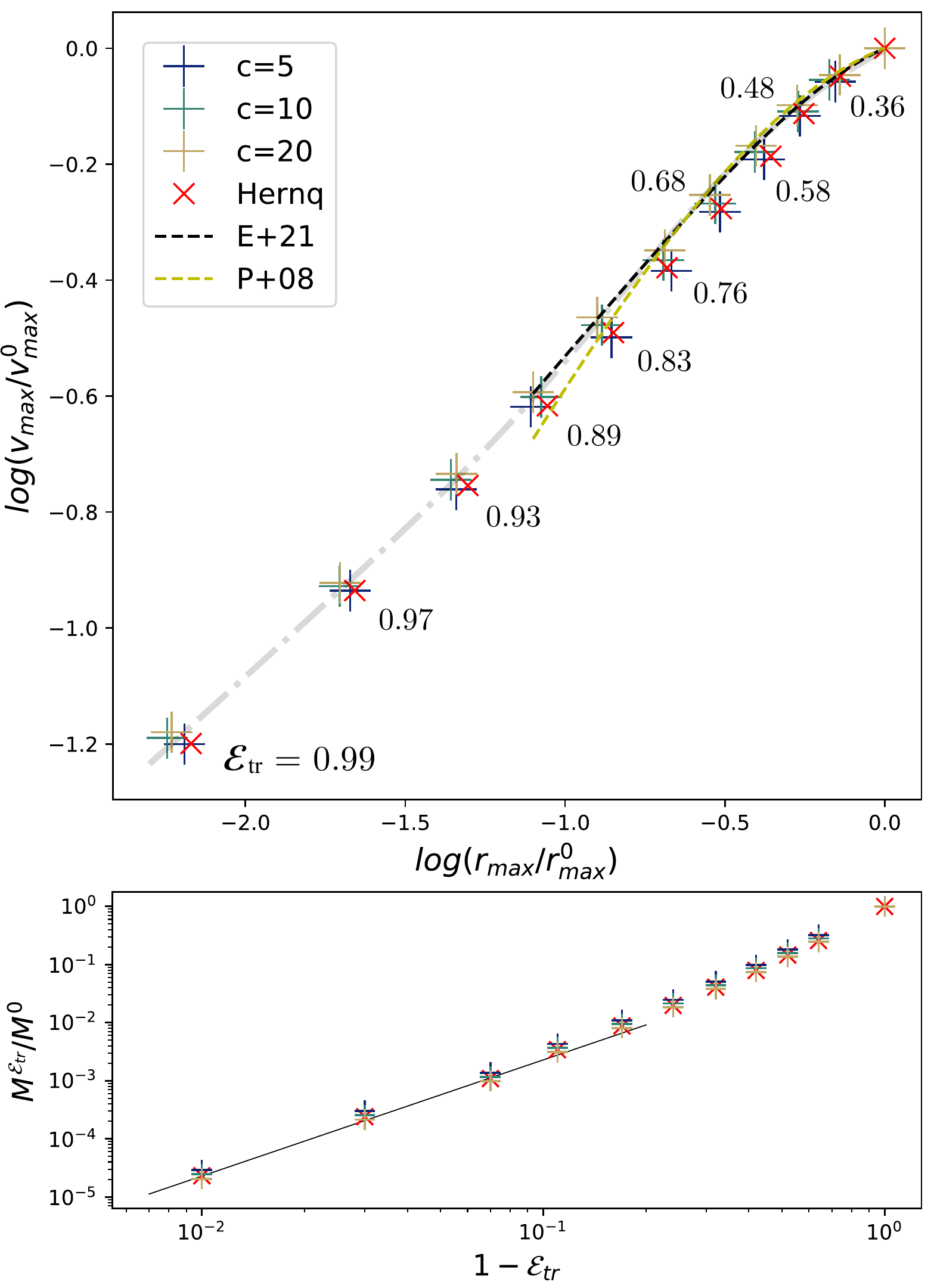}
\caption{Upper panel: the locus of re-virialised energy-truncated satellites 
in the the plane $(\rmx/\rmx^0, \vmx/\vmx^0)$, where $\vmx/\vmx^0$ is 
the ratio between the maximum circular velocity after virialization and 
the maximum circular velocity of the unperturbed, un-truncated system, and 
$\rmx/\rmx^0$ is the ratio between the location where these circular 
velocities are realised. The tidal tracks as observed in numerical 
experiments of tidal evolution of NFW satellites are shown by a gold 
\citep{Pen08} and a black dashed line (EN21). 
Points show the structural properties of energy-truncated systems
with different values of the energy truncation $\nEt$, after the 
re-virialization process. The dash-dotted 
grey line reproduces our results and is described in appendix A.
Lower panel: the fraction of mass mass enclosed in the energy-truncated satellites as a function of the energy truncation. The black line shows the asymptotic relationship for $\nEt\to 1$, and is obtained using
equation~(\ref{rhoasyf2}).}
\label{fig2}
\end{figure}
\begin{figure*}
\centering
\includegraphics[width=\textwidth]{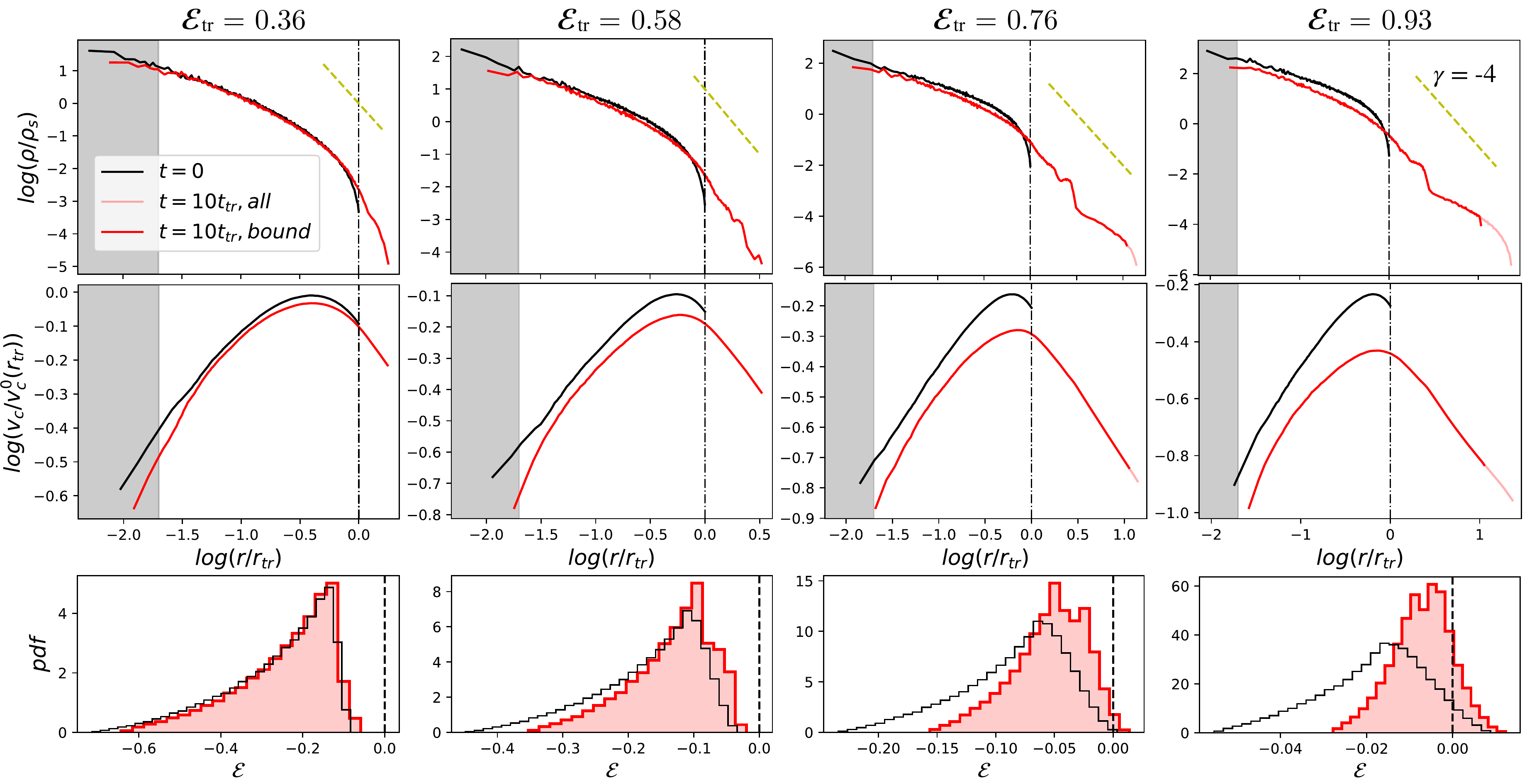}
\caption{Top (middle) panels: the density (circular velocity) 
profiles of the truncated satellites. 
Black lines indicate all material at $t=0$. Red lines represent 
self-bound material at $t=10 t_{\rm tr}$. Light-red lines include 
both bound and unbound material. The yellow dashed lines are guiding lines 
for the power-law slope $d\ln\rho/d\ln r=-4$. The vertical dot-dashed lines display the truncation radius $\rtr$. 
Bottom panels: black (red) histograms show the normalised energy distributions at $t=0$ ($t=10 t_{\rm tr}$). All distributions include
both bound and unbound material.}
\label{structexample}
\end{figure*}
\subsection{The relevance of tidal heating I}
The identity between the locus of our re-virialized truncated satellites and 
the tidal track is an extremely surprising result, and appears to highlight an unexpected 
simplicity of the tidal evolution process, at least for systems with centrally 
divergent density profiles. 
It is astonishing that our calculations entirely ignore tidal heating, and 
still reproduce the results of the full tidal evolution process!
Despite so, our results do show that the inner bound regions of the systems simulated 
by \citet{Pen08, Og19} and EN21 do not absorb appreciable amounts of energy
as a consequence of tidal heating: their structural properties are reproduced
by truncated satellites that experience no energy injection. 

In fact, this allows us to extend such a surprising result to satellites with any initial mass. For 
cosmologically motivated CDM haloes, the characteristic radius $\rs$ scales 
like $\rs\sim M_{0}^{1/3}$ down to planetary masses, or to the truncation due to free streaming \citep{Wang20}. As a consequence, 
the ratio between the energy injected by tidal heating, $E_{\rm th}$, and the 
binding energy of the undisturbed satellite at accretion, $U_0$, is independent of the initial satellite mass:
\begin{equation}
{E_{\rm th}\over U_0}\sim {M_0\rs^2\over{M_0^2/\rs}}\sim (M_0)^{0}.
\label{heatE0}
\end{equation}
This allows us to conclude that the importance of tidal heating does not increase for decreasing 
initial masses $M_0$ for cosmologically motivated systems. As the satellites simulated in the mentioned works, CDM subhalos of {\it any mass} are affected by tidal heating in similarly negligible amounts. 
This is the second key ingredient of the resilience of CDM subhaloes. 
 It should be highlighted, however, that equation~(\ref{heatE0}) does not prove that heating remains negligible
at any degree of tidal stripping. We address this in Section 3.2. 

The results above validate an approach to the process of tidal evolution based on first principles -- namely energy truncation and re-virialization -- in which 
tidal heating is entirely neglected. 
We build such a model in Section 3.
Before that, we use our numerical results to characterize the dimensionless structural properties of the re-virialised truncated satellites.

\begin{figure*}
\centering
\includegraphics[width=\textwidth]{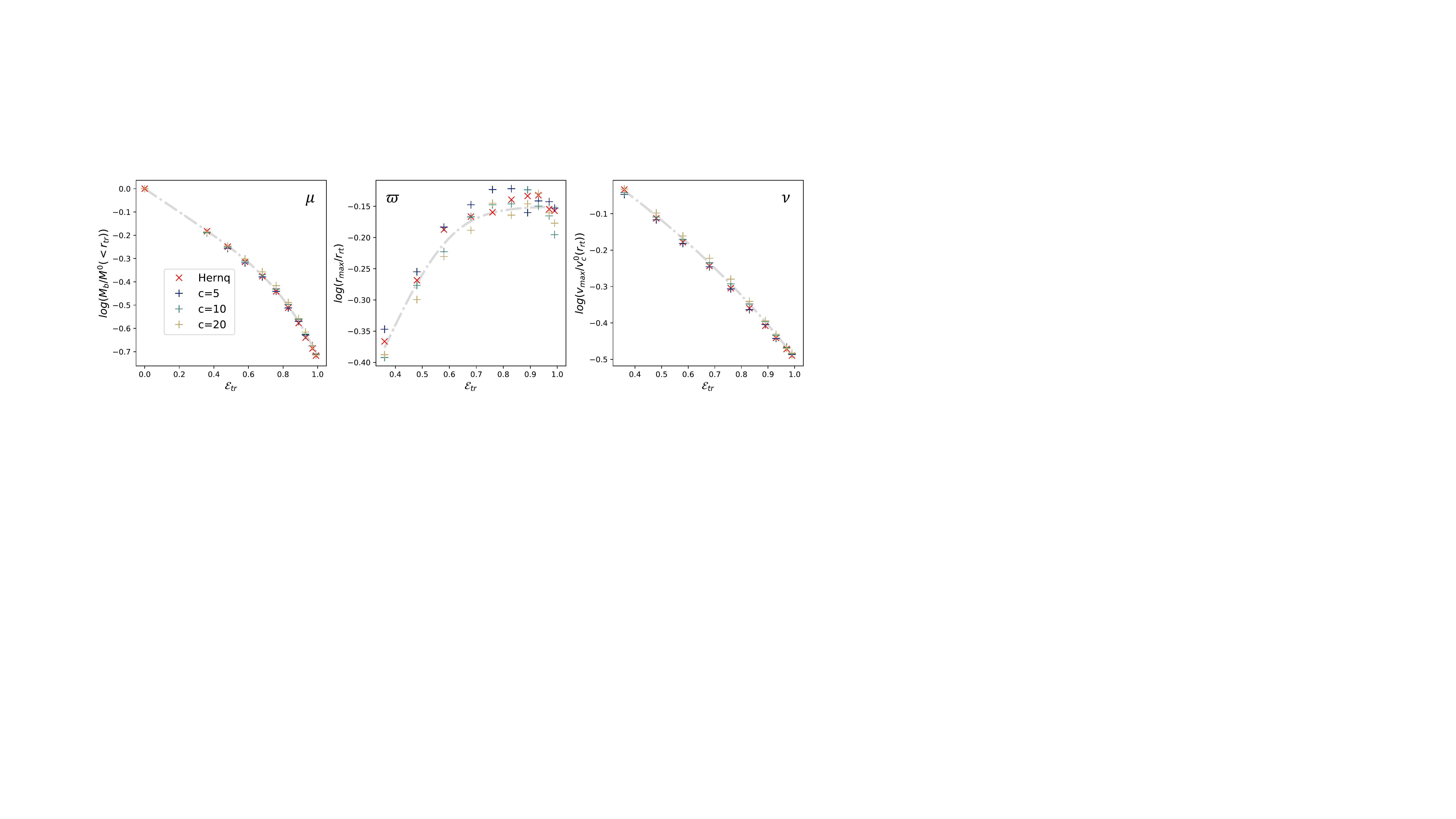}
\caption{The dimensionless structural properties of energy-truncated satellites
as a function of the truncation energy.
The left panel shows the ratio between the bound mass $M_{\rm b}$ and the initial
total mass. The middle panel shows the ratio between $\rmx$ and the truncation 
radius $\rtr$. The right panel shows the ratio between $\vmx$ and the circular
velocity at the truncation radius in the unperturbed, un-truncated satellite, $v_c^0(\rtr)$.}
\label{structexample}
\end{figure*}

\subsection{The internal structure of the re-virialized satellites}\label{numres2}

Figure~3 shows  examples of the density profiles (top panels), circular velocity
profiles (middle panels) and energy distributions (bottom panels) of our energy truncated
satellites. The different columns display different values of the truncation energy
$\nEt$, all plots pertain to an initial NFW profile with $c=20$, results are analogous for the other 
values of the concentration. Black curves refer 
to the system at $t=0$, red curves show the re-virialized systems at $t=10 t_{\rm tr}$. 
In the upper panels, lighter red curves include material that is not self-bound
as a consequence of the re-virialization. Radii are shown in units of the truncation
radius $\rtr$, circular velocities are displayed in units of the circular velocity 
at the truncation radius in the unperturbed, un-truncated satellite, $v_c^{0}(\rtr)$. 

Higher values of the truncation energy $\nEt$ correspond to more significant 
departures from virial equilibrium. This causes larger evolution as a result 
of the re-virialization itself. In fact, as shown by the energy distributions in 
the bottom row, part 
of the material in the satellite is unbound as a result of the energy truncation, 
even before the re-virialization process itself. This is the case for $\nEt=0.76$ and higher values of the truncation energy. The red histograms show that 
more material becomes unbound as the satellites
relax and expand\footnote{With a slight abuse of notation, in the bottom row of Fig.~3 we define the normalized energy,
$\nE$, as the total energy of the satellites'
particles in their own potential, both bound and unbound, normalised 
by the value of the central potential in the un-truncated density profile,
$\Phi_0$.}. 
As a consequence, the value $\nEt\approx0.76$ sets a divide in the properties of the 
density distribution of the re-virialized satellites. At lower truncation energies
satellites have sharply truncated density profiles, as their energy distributions do not extend to $\nE\to0$. 
At higher truncation energies, the presence of a population of particles
at $\nE\to0$ results in density profiles with outer power-law slopes of $d\log \rho/d\log r\approx-4$ 
\citep[e.g.,][]{Jaffe87}, as shown by the yellow guiding lines in the top row
panels. In the presence of such diffuse `haloes', dynamical timescales become very long for 
the loosely bound material at large radii, which causes the outer wings of the 
density profiles to remain partially not phase mixed at the end of our simulations. 
This does not appreciably affect the re-virialization in the central regions. 

In all cases, the density profile at the center appears to retain a power-law slope of $\gamma=-1$, 
as seen during tidal evolution in many numerical works before. The density normalization decreases,
and so does the circular velocity profile in the center. As a result of the re-virialization process, the satellite expands, increasingly so for higher truncation energies. In particular, this results in a gradual increase of the ratio $\rmx/\rtr$ for increasing values of $\nEt$.

Figure~4 collects all dimensionless structural properties of our re-virialized 
satellites, as a function of the truncation energy.
The displayed quantities represent the set of dimensionless figures that 
we could not have determined analytically without making some combination 
of assumptions on the re-virialization process itself, and that therefore we determine
numerically. In detail, these are the ratios
\begin{gather}  
  \mu \equiv \Mb/M_0(<\rtr) \label{massq} \\  
  \varpi \equiv \rmx/\rtr \label{radq}  \\
  \nu \equiv \vmx/v^0_c(\rtr) \label{velq} .
\end{gather}  
Here, $\Mb$ is the bound mass after re-virialization and $M_0(<\rtr)$ is the mass within the truncation radius in the un-perturbed, un-truncated density profile. The dimensionless
quantity $\mu$ is displayed in the left panel of Fig.~4. The middle panel shows the quantity
$\varpi$, the location of the maximum circular velocity in terms
of the truncation radius. The right panel shows $\nu$, the maximum circular velocity itself in terms of the value of the circular velocity
at the truncation radius in the un-truncated density profile, $v^0_c(\rtr)$.

It appears that the values of 
$\mu$, $\varpi$ and $\nu$ are not strongly affected by the precise value of the concentration $c$. In fact, values of the same quantities pertaining the Hernquist profile are seen to approximately follow the relations valid for the NFW density profile. In the case of the quantity $\varpi$,
we observe differences of the order of 10\% between the different models. We interpret these as scatter, most likely associated to the numerical precision of our determination of the location of the maximum circular velocity. In Section~\ref{modelasy}, we show that the similarity
of $\mu$, $\varpi$ and $\nu$ for different initial density profiles is justified, and that it is associated with the identical asymptotic divergence of the systems' distribution functions, reported in equation~(\ref{asymptf}).

It should be stressed that, for a given initial density profile $\rhsat(r)$,
the combination of equation~(\ref{rtrunc}) and the dimensionless quantities in Figure~4 are all that is needed to calculate the structural properties $(\rmx, \vmx)$ of a  satellite that re-virialises after an arbitrary energy truncation $\nEt$. We also report the relation between the truncation radius $\rtr/\rs$ and the bound mass fraction $\Mb/M_0$ in Figure~5. For convenience, we collect some simple 
parametric functions that reproduce the behaviours of $\mu(\nE)$, $\varpi(\nE)$ and $\nu(\nE)$ in Appendix A. These are shown as grey dash-dotted lines in Fig.~4.
The asymptotic behaviour of the bound mass fraction is 
also shown as as grey dash-dotted lines in Fig.~5 and 
reported in Appendix A.

\begin{figure}
\centering
\includegraphics[width=.8\columnwidth]{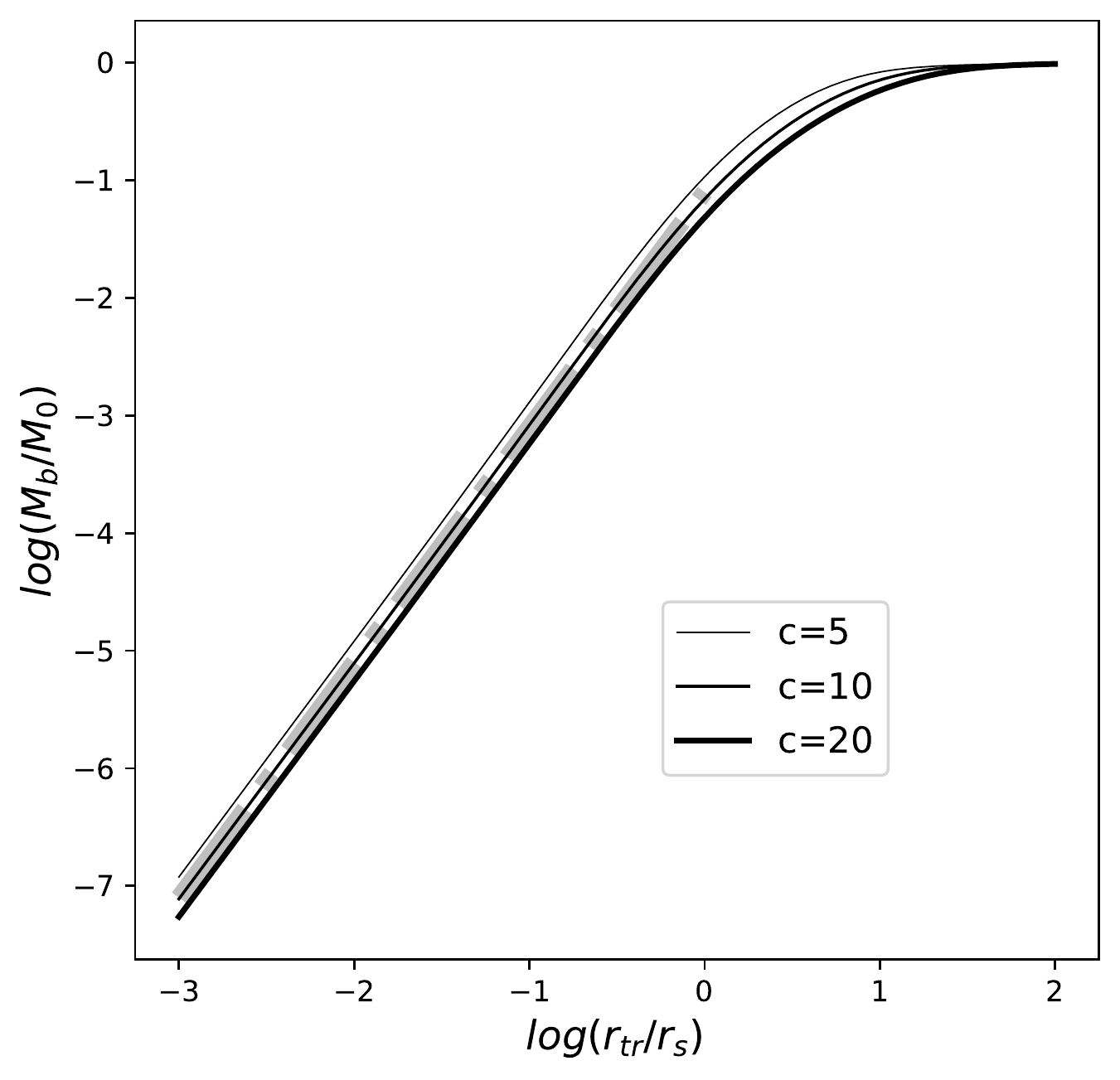}
\caption{The relation between the truncation radius $\rtr/\rs$ and the bound mass fraction $\Mb/M_0$ for 
NFW haloes with concentration $c\in\{5,10,20\}$. 
A simple parametric function valid in the asymptotic regime $\nEt\to1$ is shown as a thick dashed line and described in Appendix A.}
\label{structexample}
\end{figure}

\subsection{Truncated satellites with arbitrarily small bound mass fractions}\label{modelasy}

In the previous Sections we have shown that the tidal track is essentially 
a uni-dimensional sequence in the truncation energy $\nEt$. In the common interpretation, the tidal track is parameterized by the remnant's bound mass fraction, which in our approach is determined by the truncation energy
itself. Here, we consider the limit of `asymptotic truncation', $\nEt\to1$, in which the truncation energy approaches the bottom of the potential, causing the bound mass fraction to vanish. 

In this regime, the distribution function is fully described by its asymptotic behaviour, as in equation~(\ref{asymptf}). Similarly, the gravitational potential of the unperturbed satellite is $\Psi-1\sim (r/\rs)^{2-\gamma}$. As a consequence,
 equation~(\ref{rhotrunc}) simplifies to
\begin{equation}
\rho_{\rm sat}^{\rm asy}(x) \propto \int_{x^{2-\gamma}}^{1} t^{-{{6-\gamma}\over{2(2-\gamma)}}}(t-x^{2-\gamma})^{1\over 2}dt ,
\label{rhoasyf}
\end{equation}
where we have used $x\equiv r/\rtr$. The necessary dimensional multiplicative 
constant is set by the asymptotic condition at $x\to0$, equation~(\ref{rhotruncasy}). Analogous expressions enable the 
calculation of any other moment of the distribution function, and therefore of any kinematic property of the `asymptotically truncated' satellites.

Note that the central power-law slope $\gamma$ entirely determines the truncated satellites' properties in this regime. In particular, these are independent of the full details 
of the initial un-truncated density profile.
This justifies the similarity of the numerical results recorded
in Sections~\ref{numres1} and \ref{numres2} for different values of 
the NFW concentration,
as well as between NFW and Hernquist density profiles. 
Furthermore, equation~(\ref{rhoasyf}) shows that all satellites in the 
regime of asymptotic truncation $\nEt\to1$ have identical dimensionless 
structural properties: 
\begin{equation}
\rho_{\rm sat}^{\rm asy}(r)=\rhsat(r=\rtr)\Tilde{\rho}(x) ,
\label{rhoasyf1}
\end{equation}
where the function $\Tilde{\rho}$ does not depend on the specific value of the 
truncation energy $\nEt$. The function $\Tilde{\rho}(x)$ is unique
for all sufficiently small values of the truncation radius $\rtr/\rs$. In other words, at infinitely low bound mass fractions, the dimensionless structural properties of the energy truncated satellites are the same.
For instance, in the case $\gamma=1$, equation~(\ref{rhotruncasy}) simplifies to  
\begin{equation}
\Tilde{\rho}(x)={{\left(1-x\right)^{-3/2}}\over{x}}.
\label{rhoasyf2}
\end{equation}

The identity of the dimensionless structural properties before re-virialization 
implies their identity after the virialization process is complete. This means that the limits of the quantites $\mu$, $\varpi$ and
$\nu$ as $\nEt\to1$ are well-defined.
These values describe the re-virialization 
of all energy truncated satellites with 
sufficiently small values of the truncation radius,
despite their different bound mass fractions.
In practice, this means that it is sufficient to run a single numerical simulation to capture the 
re-virialization and subsequent structural properties at arbitrarily low bound mass fractions.
We do so for the case $\gamma=1$ using a controlled simulation analogous
to those described in Section~\ref{numres1}. We find 
\begin{equation}
\left\{\mu(1),\varpi(1),\nu(1)\right\}\approx\left\{10^{-0.75}, 10^{-0.17}, 10^{-0.50}\right\}.
\label{valsasy}
\end{equation}
In particular, we find that, even in the ultimate case of an asymptotic truncation, $\nEt\to1$, 
the fraction of mass that remains bound to the satellite as a 
consequence of the re-virialization process is 
\begin{equation}
{\Mb\over M^{\nEt}}(1) \approx 0.75.
\label{valsasy1}
\end{equation}
This proves that it is possible
to truncate an isotropic and centrally divergent density profile with $\gamma=1$
at arbitrarily high truncation energies. The re-virialized structure will be self-bound, independently of its small mass. Certainly, steeper cusps will 
behave similarly. In Section~4 we address the case
of shallower cusps $0<\gamma<1$.

Finally, it is worth noticing that 
the identity of the structural properties of the re-virialized satellites forces their locus in the plane $(\rmx/\rmx^0, \vmx/\vmx^0)$ to scale 
like $\vmx/\vmx^0\sim (\rmx/\rmx^0)^{1/2}$ at low values of $\rmx/\rmx^0$.
This is a direct consequence of the value of $\gamma$: since $M\sim r^{3-\gamma}$ at the center, we have that
\begin{equation}
\vmx\sim v_c^0(\rtr)\sim \left({{M_0(<\rtr)}\over{\rtr}}\right)^{1\over 2} \sim {\rtr}^{1\over 2}\sim {\rmx}^{1\over 2}.
\label{onehalf}
\end{equation}
%

\section{A simple model of tidal evolution}

Having described the properties of satellites truncated at arbitrary 
values of $\nEt$ and subsequently re-virialized, we look at connecting
the truncation energy with the properties of the tidal field. 
We consider a subhalo with an NFW density profile at accretion, with initial properties $\rs$, $\rhs$, 
and total mass $M_0$. For the moment, we assume that the 
subhalo evolves towards some long-lived state characterized
by the structural properties $(\rmx, \vmx)$, which allow it to 
orbit within the tidal field of its host without experiencing significant 
additional evolution. For example, the numerical results of EN21 
suggest that the structural evolution induced by the tidal 
process appears to converge on a remnant whose properties 
are uniquely determined by the host's tidal field at pericenter.

The classical definition of the tidal radius stems from the analysis 
of the forces in the vicinity of the satellite's centre.
At the tidal radius\footnote{Notice that we use different symbols for the truncation radius, $\rtr$, and the tidal radius 
$\rt$, as these are {\it not} the same quantity in our model.} $\rt$ the gravitational attraction of the satellite
itself and the forces from the main host are balanced:
\begin{equation}
\left[-{d\over{dr}}\Phi_{\rm sat}(r)+\lambda^{\rm p}_1 r\right]_{\rt}
= -{{G M_{\rm sat}(\rt)}\over{\rt^2}}+\lambda_1^{\rm p} \rt
=0 .
\label{tidalr}
\end{equation}
Here, $\lambda_1$ is the largest eigenvalue of the effective tidal tensor
\citep[e.g.,][]{Renaud11}, and $\lambda_1^{\rm p}$ is its value at the pericentric radius, $\rp$. 

Our results of Section~2 motivate using
equation~(\ref{tidalr}) as an equation for the value of the truncation energy itself, assuming that the properties of the tidally evolved satellites are well
described by the uni-dimensional sequence of truncation energies. 
We can expect such an approach to return the correct
dimensional scalings of the structural properties of tidally evolved remnants.
In turn, it is unclear to which accuracy this approach can return dimensionless coefficients. First, as shown by EN21 and previous works, tidally evolved satellites appear
to have exponentially truncated density profiles for the bound material. This is not the case for the density profiles that result from re-virialization, which are charaterized by $\rho\sim r^{-4}$ at large radii in the case of significant mass loss. Second, our model adopts a sharp energy truncation, while 
numerical results suggest a more gentle transition and secondary dependences \citep[e.g.,][]{Read06, Drakos20}.
This may suggest that the 
values of the structural parameters collected in Section~2.5 may not fully capture the 
dimensionless structural properties of stripped satellites. We stress, however, that this
is an issue of coefficients rather than dimensional scalings. 
As long as the energy distribution is unaffected below some energy value our dimensional scalings are still valid.  Therefore,
we proceed using the same set of symbols defined in Section~2.

\subsection{Subhaloes in a tidal field}

We focus on the regime of important mass loss $\nEt\to1$, or equivalently $\rtr/\rs\to0$. 
We re-write $\rt=\xi\rtr$, where $\xi$ is a number, and, for brevity, we are assuming 
$\xi=\xi(\nEt=1)$ -- we drop the explicit dependence on the truncation energy 
from now on. Analogously, we highlight the bound mass of the re-virialized satellite by rewriting $M_{\rm sat}(<\xi\rtr)=\zeta M_{\rm b}$.
With these substitutions, and using that, for $\gamma=1$, 
\begin{equation}
M_{\rm b}=2\pi\mu\rhs \rs\rtr^2 ,
\label{masssmall}
\end{equation} 
equation~(\ref{tidalr}) gives us
\begin{equation}
\rtr = \rs{{G \rhs}\over{\l1}}2\pi\mu{{\zeta}\over{\xi^3}}.
\label{tidalr1}
\end{equation}
This equation determines the energy truncation
-- i.e. the truncation radius $\rtr$ -- which corresponds to the 
re-virialized satellite having a tidal radius $\rt$ of $\rt=\xi\rtr$. 
Under the assumption that the coefficients of
equation~(\ref{valsasy}) are approximately valid here, the dimensionless quantity $\zeta/\xi^3$ is the only unknown in this equation. The 
initial properties of the satellite $\rhs$ and $\rs$ are fixed, and so is the host's tidal field at pericenter.  
In terms of physical mechanisms, it seems plausible to imagine that the satellite 
experiences a progressive energy truncation in the form of tidal stripping. This essentially stops when the truncation radius 
$\rtr$ is such that the majority of the re-virialized structure 
sits within the remnant's tidal radius $\rt$.
In other words, once $\xi^3/\zeta$ is large enough, the tidal stripping and the 
subsequent evolution slow down considerably, and a long lived state is achieved. 
This has the structural properties:
\begin{gather}
\rmx = \varpi \rtr = \rs{{G \rhs}\over{\l1}}2\pi\mu\varpi{{\zeta}\over{\xi^3}},\label{rmrtr}\\
\vmx = \nu v_c^0(\rtr) = \vs {\nu\over\sqrt2}\left({\rtr\over\rs}\right)^{1\over2} \label{vmrtr},
\end{gather}
where we have defined $v_{\rm s}\equiv\sqrt{4\pi G\rhs \rs^2}$. Note that the left-hand sides of these equations describe the tidal track in the limit of small characteristic radii and velocities, parametrized by the value of the truncation radius $\rtr$. 
This model then suggest a process of tidal evolution which starts with an initial fast phase of inequilibrium, in which material is quickly lost because the initial density profile is too extended with respect to its own tidal radius, and proceeds in a progressively slower manner, as the system approaches a state in which it is almost fully contained within its own tidal radius.

It is well known that a tidal truncation as in  equation~(\ref{tidalr}) is equivalent to requiring that the mean density of the remnant and the mean density of the host within the pericenter are multiples of each other,
with a coefficient that depends of the host's density profile
and any assumptions on the remnant's orbit \citep[e.g.,][]{Taylor01,Drakos20}. This is also equivalent to requiring that  
the characteristic dynamical timescale of the remnant, $\Tmx\equiv\rmx/\vmx$, is a fixed fraction of the dynamical timescale of the host at the pericentric radius, $T_{\rm host}\equiv r_{\rm p}/V_c(r_{\rm p})$, where $V_c(r_{\rm p})$
is the circular velocity of the host at the pericentric radius.
This fact has recently been highlighted in numerical simulations by EN21. In the context of our simple model, we find that 
\begin{equation}
{\Tmx}= {\rmx\over\vmx} = {\varpi\over\nu}{\rtr\over{v_c^0(\rtr)}} = {\varpi\over{\sqrt{2\pi}\nu}}\left({\rtr\over{G\rhs\rs}}\right)^{1\over 2},
\label{sattime}
\end{equation}
which, using 
equation~(\ref{tidalr1}) can be written as
\begin{equation}
{\Tmx}= {{\varpi\mu^{1\over2}}\over{\nu}}\left({{\zeta}\over{\xi^3}}\right)^{1\over 2}(\l1)^{-{1\over 2}} .
\label{sattime2}
\end{equation}
In the case of a 
self-similar density profile like used by EN21 in their simulations, or
in the more general case in which the pericentric radius lies 
considerably within the host's scale radius, $(\l1)^{-{1/2}}\propto T_{\rm host}$. 

It is useful to consider explicitly the case of an NFW host halo,
with characteristic radius $\rh$ and characteristic density $\rhh$.
For orbits that result in strong tidal fields, $\rp\ll\rh$, the 
largest eigenvalue of the tidal tensor is 
\begin{equation}
\l1 = {2\pi}G\rhh{\rh\over \rp}.
\label{masssmall}
\end{equation}
This gives us 
\begin{gather}
\rmx = \rs{{\rp\rhs}\over{\rh\rhh}}\mu\varpi{\zeta\over \xi^3},
\label{rmxNFW}\\
\vmx = v_{\rm s} \left({\rhs\rp\over\rhh \rh}\right)^{1\over 2}\nu\left({\mu\over2}{\zeta\over \xi^3}\right)^{1\over 2}.
\label{vmxNFW}
\end{gather}
These provide us with the scalings for the structural properties 
of CDM subhaloes orbiting the innermost regions of an NFW host 
in the case of significant mass loss. Equations~(\ref{rmxNFW}) 
and~(\ref{vmxNFW}) are valid for any initial subhalo mass $M_0$
and arbitrarily low bound mass fractions. The basic hypothesis behind these equations is that tidal heating is negligible. 

To complete the model, we require a determination of the value $\zeta/\xi^3$. Other numerical coefficients in equations~(\ref{rmxNFW}) 
and~(\ref{vmxNFW}) are estimated in Section~2.5. 
Since $\rmx$ and $\rtr$ have similar magnitudes, it seems reasonable to argue that $\xi$ has to be 
of the order of a few if most of the remnant itself is to be contained within the tidal radius.
However, the only way to provide a reliable determination of the value
$\zeta/\xi^3$ is through a numerical simulation. For this purpose, 
we consider the results of EN21, who find that $\Tmx \approx T_{\rm host}/4$ once tidal evolution has slowed down substantially. Using equation~(\ref{sattime2}),
this gives us\footnote{In passing, we note that, assuming $\zeta\approx1$, this give us a value of $\xi\approx 2$, in line with our naive expectation.} $\zeta/\xi^3\approx\nu^2/8\varpi^{2}\mu$.  In conclusion, for an NFW host, we have
\begin{gather}
\rmx \approx \rs{{\rp\rhs}\over{\rh\rhh}}{\nu^2\over {8\varpi}}\approx 1.8\times10^{-2}\rs{{\rp\rhs}\over{\rh\rhh}},
\label{rmxNFW1}\\
\vmx \approx \vs\left({\rhs\rp\over\rhh \rh}\right)^{1\over 2}{\nu^2\over 4\varpi}\approx 3.7\times10^{-2} v_{\rm s} \left({\rhs\rp\over\rhh \rh}\right)^{1\over 2},
\label{vmxNFW1}\\
\Tmx\approx {T_{\rm host}\over 2\sqrt2},\label{TmxNFW1}
\end{gather}
We test these dimensional scalings and coefficients in Section~3.3. First, however, we return on the issue of tidal heating.

\subsection{The relevance of tidal heating II}

Equation~(\ref{heatE0}) shows that the importance of tidal heating does not increase for subhalos with lower mass at accretion. Therefore, the combination of equation~(\ref{heatE0}) and Figure~2
have allowed us to conclude that tidal heating is essentially negligible in the initial phases of the tidal evolution of subhaloes of any mass at accretion. 
Equation~(\ref{heatE0}), however, does not guarantee that tidal heating 
remains negligible once tidal evolution
has substantially affected the structural properties of the subhalo itself, and made them depart from the properties of CDM haloes. The evolution may, in principle, proceed towards
structural configurations of the remnant in which tidal heating is important.

\begin{figure*}
\centering
\includegraphics[width=.9\textwidth]{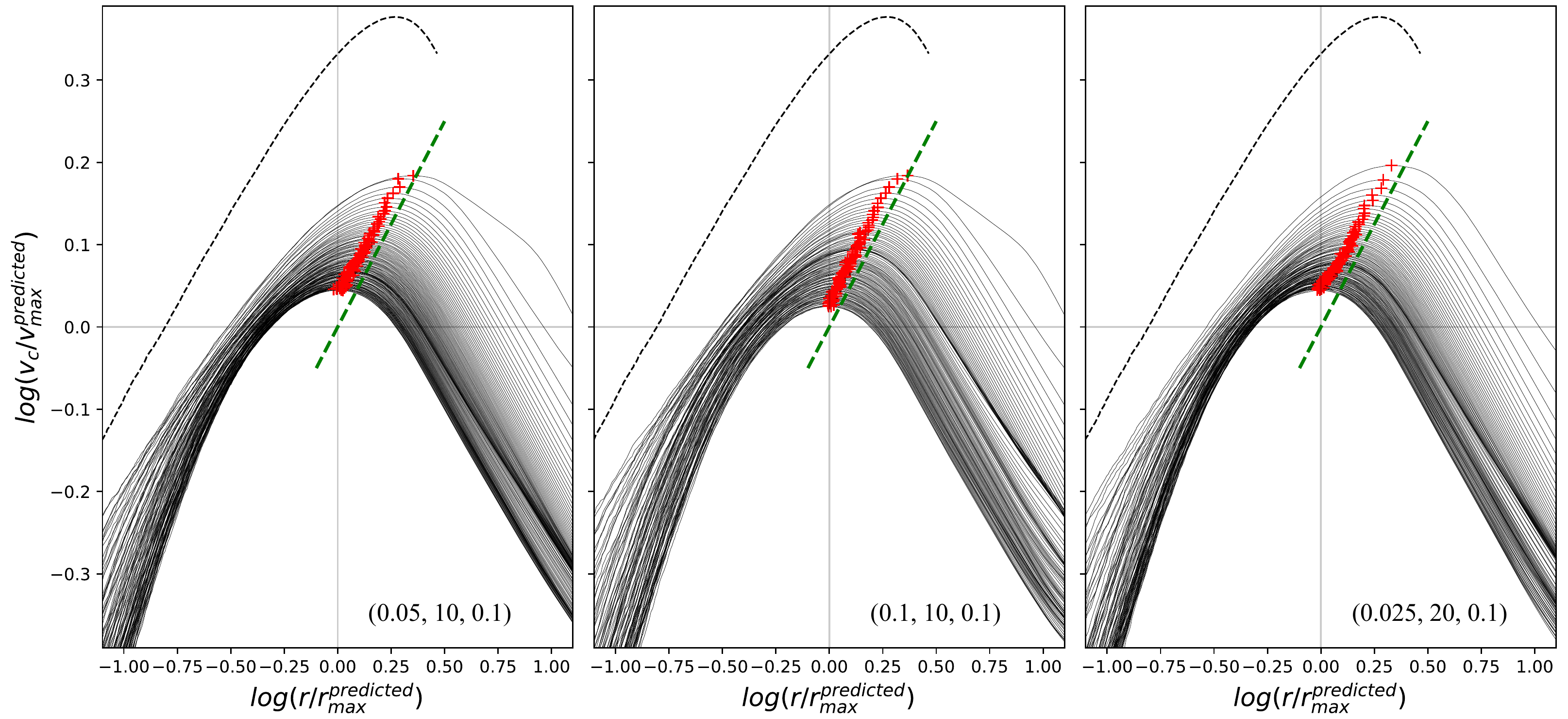}
\caption{Numerical tests of the model. The three panels show the 
circular velocity profiles of three different tidal evolution experiments, corresponding to combinations of subhalos/host properties in the regime of extreme mass loss, $\Mb/M_0\approx10^{-4.5}$. The radius and circular velocity values are scaled
by the predictions of our model. The different curves span a time
interval of $\approx170$ dynamical timescales of the remnant.
The green dashed line shows the locus of re-virialized truncated satellites in the regime of extreme mass loss, as predicted by equations~(\ref{rmrtr}) and~(\ref{vmrtr}). }
\label{structexample}
\end{figure*}

Numerical results seem to suggest that that
is not the case: numerically determined tidal tracks coincide with the locus 
of re-virialised energy-truncated satellites down to $\rmx/\rmx^0\approx0.1$, This
shows that tidal heating remains of little importance at least while $\nEt\lesssim0.9$, 
or equivalently, for $\Mb/M_0\gtrsim 10^{-3}$.  
In fact, we can settle this for all degrees of mass loss using our model.

Our model shows that tidal heating remains unimportant at all values of the bound mass fraction: evolution along the sequence 
of re-virialized energy truncated remnants does not cause the
importance of tidal heating to increase with decreasing
bound fractions. That is because, for $\nEt\to1$, we have that $\Mb\sim\rmx^2$. Therefore, we find that
\begin{equation}
{E_{\rm th}\over U}\sim {\Mb\rmx^2\over{\Mb^2/\rmx}}\sim \Mb^{1\over2}.
\label{heatE1}
\end{equation}
This indicates that heavily stripped satellites are in fact
{\it less} susceptible to tidal heating then their progenitors, and that the stripping process itself causes tidal heating to become progressively less important. 
Since the 
long lived states described by equations~(\ref{rmxNFW1}) and~(\ref{vmxNFW1})
are obtained under the hypothesis that tidal heating is always negligible, equation~(\ref{heatE1}) guarantees
that our model is consistent. Furthermore, it ensures that, once the remnant is described by equations~(\ref{rmxNFW1}) and~(\ref{vmxNFW1}), it does not evolve further due to heating.
Equation~(\ref{heatE1}) is the third and last  
ingredient in the tidal resilience of CDM subhaloes. 

We should clarify that equations~(\ref{heatE0}) and~(\ref{heatE1}) do not guarantee that tidal heating is uniformly negligible for all CDM subhaloes, in all tidal fields. Our anchor points are the numerical results of \citet{Pen08, Og19} and EN21, which have analysed cosmologically motivated haloes orbiting NFW or 
$\rho\sim r^{-2}$ hosts. First, deviations from the mean mass-concentration relation in the subhaloes at accretion certainly 
affect the value of the ratio in equations~(\ref{heatE0}) and~(\ref{heatE1}) by some dimensionless factor. It is therefore possible that tidal heating may become important for significantly under-concentrated CDM subhaloes \citep[see e.g.,][]{Amo19}. The combination of
equations~(\ref{heatE0}) and~(\ref{heatE1}) suggests that this may be the case during the initial phases of the tidal evolution process. Second, the presence of a massive disk may also result in a substantial increase of the energy injected in the remnants by tidal shocks \citep[][]{Pen10,Donghia10,Kelley19}. This is not included in our model and further numerical study is required to ascertain whether this invalidates our basic hypothesis.

\subsection{A numerical test}

We test the scalings and coefficients of equations~(\ref{rmxNFW1}) and~(\ref{vmxNFW1}) 
by simulating the tidal evolution of three highly stripped subhaloes orbiting the inner regions of NFW hosts, with different combinations of properties.
We choose a fiducial model in which:
\begin{equation}
\left({\rs\over \rh}, {\rhs\over \rhh}, {\rp\over \rhh}\right)=\left({1\over 20},{10},{1\over10}\right) .
\label{fiducial}
\end{equation}
Two additional models are chosen so that the same set of parameters has
values $(1/10,10,1/10)$ and $(1/40,20,1/10)$. Such a set allows us to test the dimensional scalings derived in Section~3.1.

For this set of properties, our model predicts bound mass fractions of the order of $\Mb/M_0\approx10^{-4.5}$. This makes it challenging to follow the long term evolution of these systems numerically. On the one hand,
preserving the cusp requires high force resolution, i.e. small softening lengths. On the other, this implies that the remnant has to be resolved with large numbers of particles in order to suppress relaxation processes, which, in the long term, also tend to erase the cusp. For example, in order to resolve the remnant with $\approx10^5$ particles, the system should be composed of $N\gtrsim10^9$
particles at accretion, which is unrealistic. Therefore, we
choose to initialize these simulations using a set of `partially truncated' satellites, i.e. with systems with a truncation energy that is lower than imposed by the tidal field. This spares us the computational cost of simulating 
the largest majority of particles that are quickly lost from the system. We stress that the satellites in these simulations experience significant 
tidal stripping during the simulations themselves:
the final state of the remnants is not determined by the energy truncation chosen as initial conditions, rather, it is set by the tidal field of the host. 
For definiteness, we use an $\nEt=0.97$ energy truncation,
and a parent concentration of $c=5$. We resolve the truncated satellites with $N=5\times10^5$ particles, which guarantees 
$\approx10^5$ particles in the remnants. The three subhaloes are 
evolved on circular orbits for more than 60 orbital periods, which, through equation~(\ref{TmxNFW1}), is equivalent to about 170 dynamical times of the remnants themselves. 

Figure~6 shows the results of these simulations, in terms of the 
circular velocity profile of the remnant at equally spaced time intervals, corresponding to $\approx 1.7$ dynamical times of the remnant, as quantified by equation~(\ref{TmxNFW1}). Both radius and circular velocity are normalised by the characteristic scales predicted by our model, 
equations~(\ref{rmxNFW1}) and~(\ref{vmxNFW1}). The fact that the three
panels appear essentially identical when normalised in this way shows that the dimensional 
scalings of equations~(\ref{rmxNFW1}) and~(\ref{vmxNFW1}) are correct. 

The dashed lines
show the circular velocity profiles of the energy-truncated
satellites at the beginning of the simulation.
As the simulation starts, the system re-virializes and starts to get stripped. This results in an initial phase of rapid evolution: within the first two
dynamical times, all three satellites lose a large fraction of their mass. After that, we see continuous, but increasingly slower evolution. This appears to validate the picture laid down in Section~3.1. 

The red crosses highlight the location and values
of the maximum circular velocity $\vmx$. These points are seen tracking the relationship set by
equations~(\ref{rmrtr}) and~(\ref{vmrtr}). This is highlighted
by a green dashed line, which has the characteristic slope 
captured by equation~(\ref{onehalf}). 
This confirms that our simple model captures the evolution of highly stripped subhaloes.
Furthermore, the coefficients collected in Section~2.5, and determined exclusively by re-virialization, appear to provide a good approximation here.
There appears to be a small shift between the model curve and the locus of the numerically determined points of maximum circular velocity,
of the order of $\approx-.1~$dex in $\rmx$. This may be caused by numerical effects or, as mentioned earlier, by the details 
of the actual energy truncation associated with stripping.
Similarly, the characteristic values 
$\rmx$ and $\vmx$ are not seen to reproduce the values predicted 
in equations~(\ref{rmxNFW1}) and~(\ref{vmxNFW1}) exactly. 
Rather, within the simulated time-span, we see the remnants approach such values. Differences however, are small, of the order of $0.05~$dex in $\vmx$. This also appears to confirm that, once the remnant is approximately entirely contained its tidal radius, further evolution
slows down considerably. 

We should notice that, despite the care put to adjust the numerical resolution in our experiments, our remnants appear to become affected by
spurious numerical effects towards the end of our simulations. In particular, the central regions of the circular velocity profile shows signs of progressive steepening as time increases, a clear sign of artificial core formation.
In turn, this may play some role in the fact that, while we observe
a progressively slower evolution, this is not seen to entirely stall. At the same time, the orbit of our remnants appears to decay, by about $5\%$ by the end of the simulations. This is the result of dynamical friction on the already stripped material \citep[e.g.,][]{vdb18}. Although this is a small effect, it can clearly also contribute to the the continued evolution at late times. 
In conclusion, we are unable to disentangle these small effects from the ideal, unperturbed, tidal evolution of the remnants. This means that it is not possible to ascertain 
whether the remnants would actually stop evolving when the properties predicted by 
equations~(\ref{rmxNFW1}) and~(\ref{vmxNFW1}) are reached,
or whether they would continue evolving ever more slowly
following the tidal track.

In practical terms, however, this is of little importance. Our numerical experiments evolve for a total of $\approx170$ dynamical times, or 60 orbital periods. Scaling the host to a MW-sized halo would give us a total of 12 Gyr. This is considerably longer than the number of orbital periods experienced by most cosmological subhaloes. 
Our simulated haloes do not appear to reach the values 
predicted by equations~(\ref{rmxNFW1}) and~(\ref{vmxNFW1}),
but these still provide acceptable estimates during a very long time interval. For more precise estimates, one should accompany 
our model of the tidal track with a suitable description of the the rate at which subhaloes moves along it. This is deferred to future study. 

Finally, it is interesting to notice such a long phase of slow evolution is characterized by a perhaps unexpected scaling law,
$\Mb\sim t^{-1/3}$. This is illustrated in Figure~7, which displays the product $\rmx\vmx^2$ -- in units of the value predicted by equations~(\ref{rmxNFW1}) and~(\ref{vmxNFW1}) -- 
as a function of time for our fiducial simulation. 
The black line shows the scaling $t^{-1/3}$. This is a rather slow evolution. In particular, it is slower than what one would predict with a model $d\Mb/dt \propto -\Mb/t_{\rm dyn}$ \citep[e.g.][]{Taylor01,Jiang16}. In this regime, the dynamical time of the remnant is approximately constant,
which implies such a model would predict an exponential evolution. Analogously,
this evolution is not caused by tidal heating:
use of our equation~(\ref{heatE1}) would also result in a faster 
evolution. Once again, we stress that the numerical effects and 
the self-friction mentioned above do contribute to this evolution. Therefore, our remnants would evolve even more slowly in absence of these spurious effects. 

\begin{figure}
\centering
\includegraphics[width=.85\columnwidth]{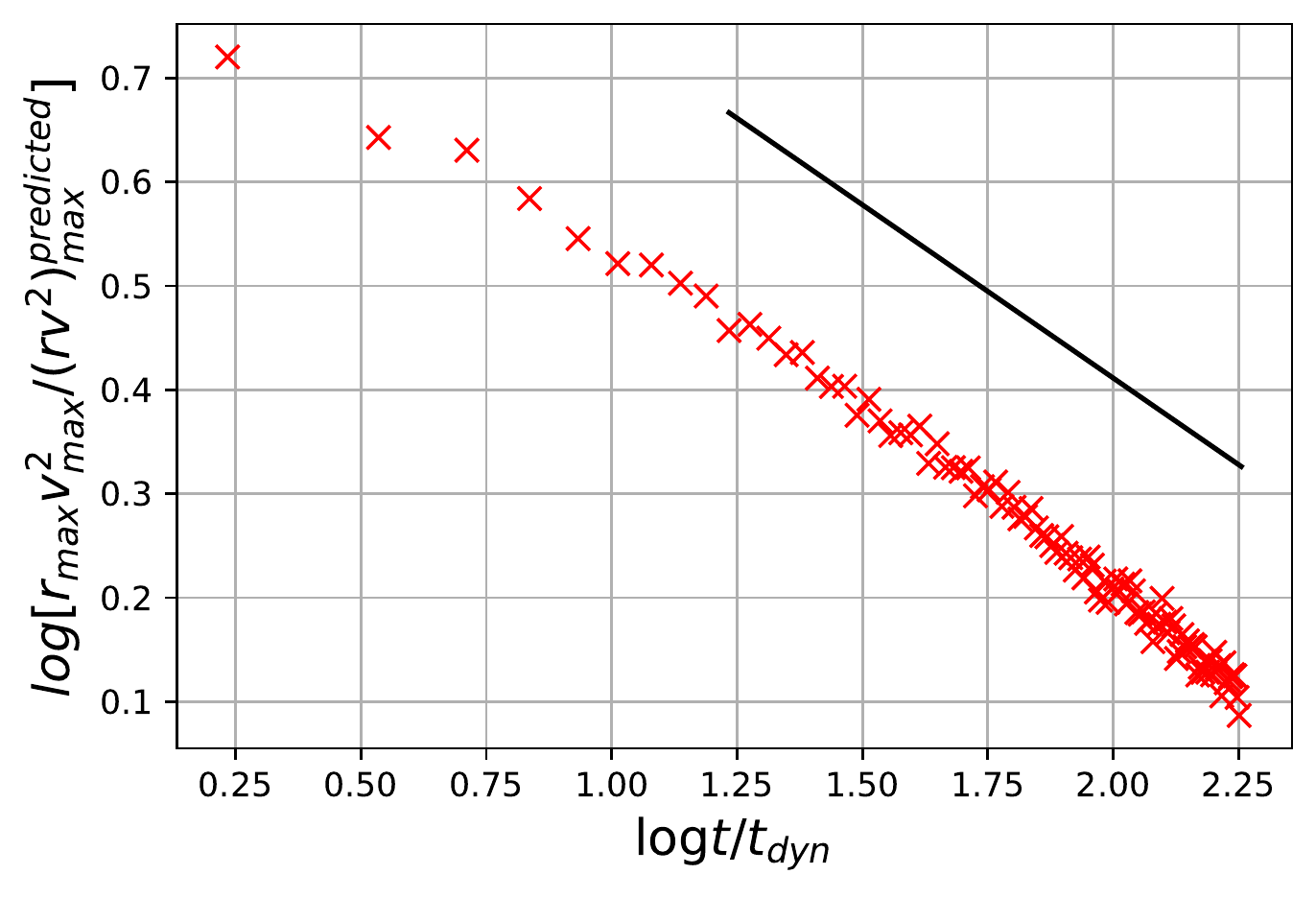}
\caption{The time dependence of the product $\rmx\vmx^2$,
proportional to the bound mass $\Mb$, as a function of time for our fiducial numerical experiment. The black line shows the scaling $\Mb\sim t^{-1/3}$.}
\label{fig1}
\end{figure}
\begin{figure}
\centering
\includegraphics[width=.85\columnwidth]{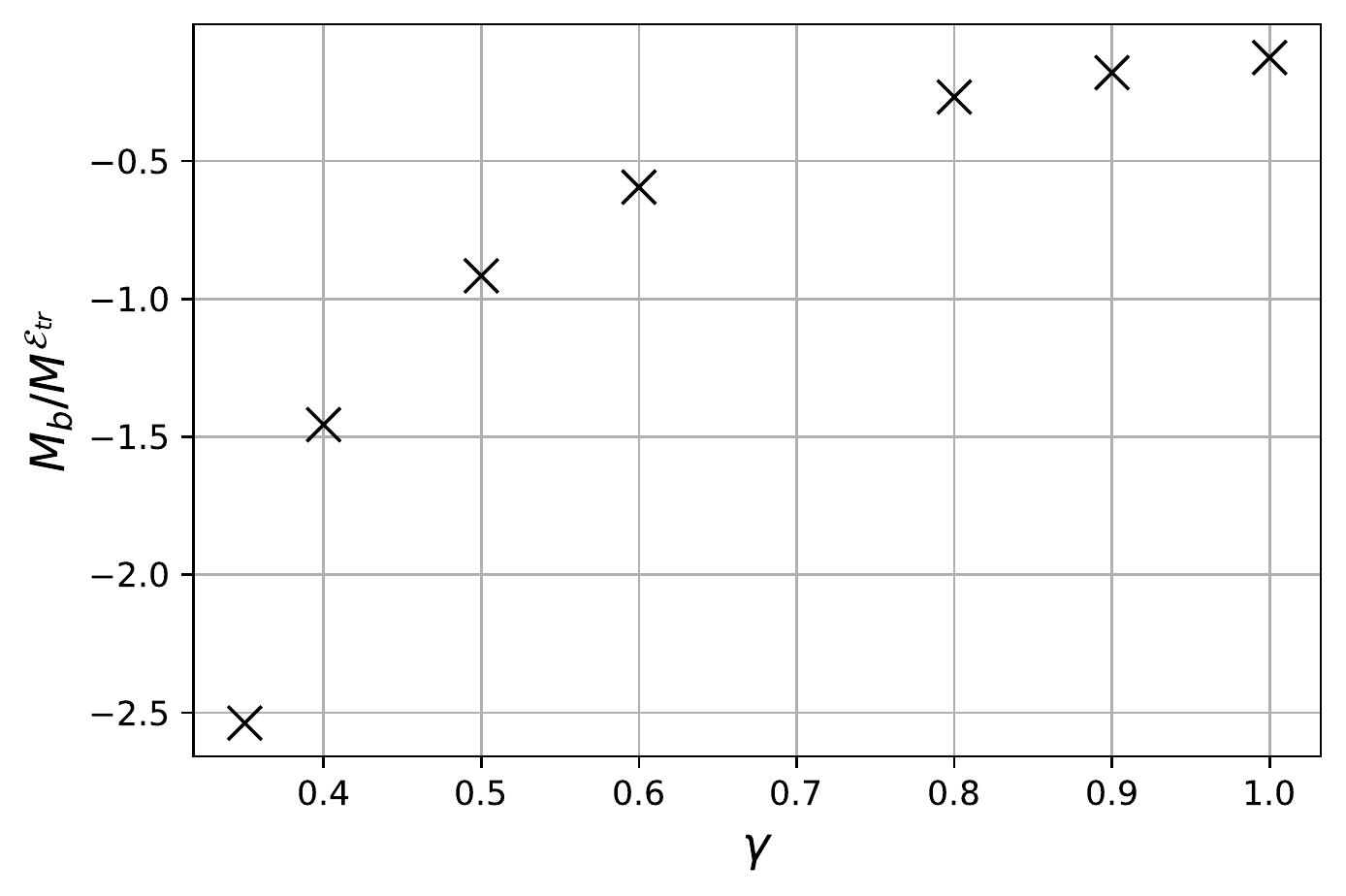}
\caption{The ratio between the bound mass after the re-virialization process, $\Mb$,
and the total mass of the energy truncated systems, $M^\nEt$, for asympotically truncated systems, $\nEt\to1$, and for 
different values of the central density slope $\gamma$.}
\label{fig1}
\end{figure}

\section{The role of the cusp index $\gamma$}

We have seen that the asymptotic condition~(\ref{rhotruncasy}) is satisfied for 
any value of the central power-law index $\gamma>0$. This does not ensure, 
however, that density profiles with cusps shallower than $\gamma=1$ can also be stripped down to arbitrarily
low bound mass fractions. First of all, that is because 
equation~(\ref{rhotruncasy}) does not guarantee that the
re-virialization process results in bound remnants for arbitrary values of the truncation energy $\nEt$.

For instance, shallower power law slopes $\gamma$ result in stronger departures from virial equilibrium at constant values of the truncation energy $\nEt$. Even before re-virialization,
we find that values $\gamma\lesssim0.5$ result in positive values of 
the total energy of the truncated satellite when $\nEt\to1$. While this does not exclude the possibility 
of a bound remnant following re-virialization \citep[see also][]{vdb18}, 
this suggests that re-virialization will cause the truncated satellite to lose a larger fraction of its mass. 

We perform a set of simulations addressing the re-virialization of 
asymptotically truncated satellites, $\nEt\to1$, with different 
values of the central power-law slope $\gamma$. These simulations are analogous to
those described in Section~2: the truncated satellites are evolved as isolated
systems for a total of $10t_{\rm tr}$. Figure~8 shows our results for the ratio 
between the mass of the bound remnant after re-virialization, $\Mb$,
and the total mass of the truncated satellite $M^\nEt$. This ratio
appears to decrease very sharply for shallower and shallower cusps. However, we detect bound re-virialized
remnants in most of our simulations, and certainly down to our mass resolution limit. Our last 
bound remnant is for the case $\gamma=0.35$, and 
comprises only $N\approx290$ particles. Therefore,
we cannot exclude that re-virialization does result in a bound remnant for even lower values of $\gamma$. 

\begin{figure*}
\centering
\includegraphics[width=.9\textwidth]{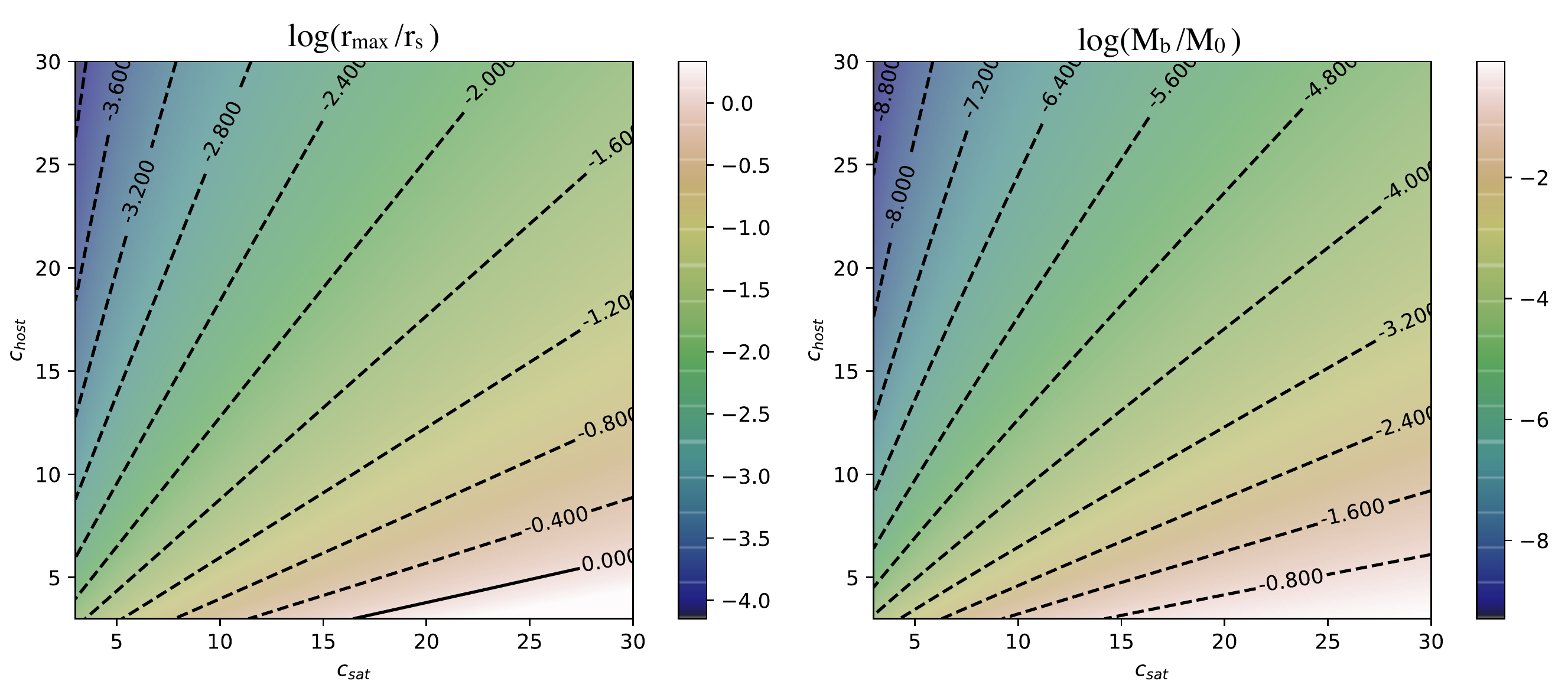}
\caption{The values of the ratios $\rmx/\rs$ and $\Mb/M_0$
predicted by our model model for subhaloes on orbits
with $\rp=\rh$, as a function of the concentrations of satellite and host.}
\label{rpeqrh}
\end{figure*}

Clearly, it still remains a question to what degree density profiles with $\gamma<1$ are affected by tidal heating.
A blind use of our model would still result in consistent predictions: using that $\Mb\sim\rtr^{3-\gamma}$, we have 
\begin{equation}
{E_{\rm th}\over U}\sim {\Mb\rmx^2\over{\Mb^2/\rmx}}\sim \rtr^{\gamma},
\label{heatE2}
\end{equation}
with a similar decreasing behaviour as in equation~(\ref{heatE1}). However, this is clearly unjustified for arbitrary values of $\gamma$.
Despite that, Figure~8 and equation~(\ref{heatE2}) do suggest that the infinite resilience to tides of centrally divergent density profiles is not singularly tied a value of power-law index of $\gamma=1$. All indications are that
this is a `continuous property', which likely extends to 
density profiles with somewhat shallower slopes. 
Nonetheless, the importance of tidal heating does increase as $\gamma$ decreases, which suggests that the fate of systems
with shallower slopes may depend on their specific properties. This, however, can not be detailed without further numerical study.

\section{Cosmological Subhaloes}

Here, we concentrate on the long lived state of  subhalos
described by equations~(\ref{rmxNFW1}) and~(\ref{vmxNFW1}), 
in the context of cosmological CDM subhalo populations.
First, note that equations~(\ref{rmxNFW1}) and~(\ref{vmxNFW1})
suggest a population of highly stripped subhaloes. To get a quick estimate, consider a typical cosmological orbit with $\rp/\rh\approx1$ \citep[e.g.,][]{Wetzel11, Jiang15}. For haloes on the mass concentration relation, the ratio $\rhs/\rhh$ at accretion is independent of the accretion redshift itself, and is a function of the concentrations of satellite and host, $c_{\rm sat}$ and $c_{\rm host}$ respectively. Assuming $\rhs/\rhh\approx 2$ we find  $\rtr/\rs\approx 5\times10^{-2}$. From Fig.~5 we see that this implies a considerable degree of mass loss, of the order of $\Mb/M_0\approx 10^{-4}$. 

Figure~9 shows a more comprehensive quantification.
The left panel displays the ratio $\rmx/\rs$ as a function of the concentration of satellite and host, assuming $\rp/\rh=1$. Through the relationships in Fig.~5, these imply the bound mass fractions $\Mb/M_0$ displayed in the right panel. Unless the 
subhalo is significantly more concentrated than the host at accretion, it will experience a considerable level of mass loss. 
Note in particular that lower satellite-to-host mass ratios at infall imply higher final 
bound mass fractions. This may appear counter intuitive, but is due to the higher concentrations of low mass CDM haloes. A lower subhalo mass at accretion 
means higher concentration, which therefore causes low-mass subhaloes to retain a larger fraction of their mass 
with respect to more massive subhaloes on the same orbit. 
It should be highlighted that the quantities displayed in Fig.~9 do not take into account either the growth of the host after accretion or the evolution of the subhalo's orbit. 
 It is well known that massive subhaloes experience significant dynamical friction and experience orbital radialization \citep{Amo17}. These will result in increased stripping. Here, however, we are mainly interested in subhaloes with low satellite-to-host mass ratios, for which dynamical friction is negligible. Similarly, we neglect the effects of the host's evolution.

\begin{figure}
\centering
\includegraphics[width=.95\columnwidth]{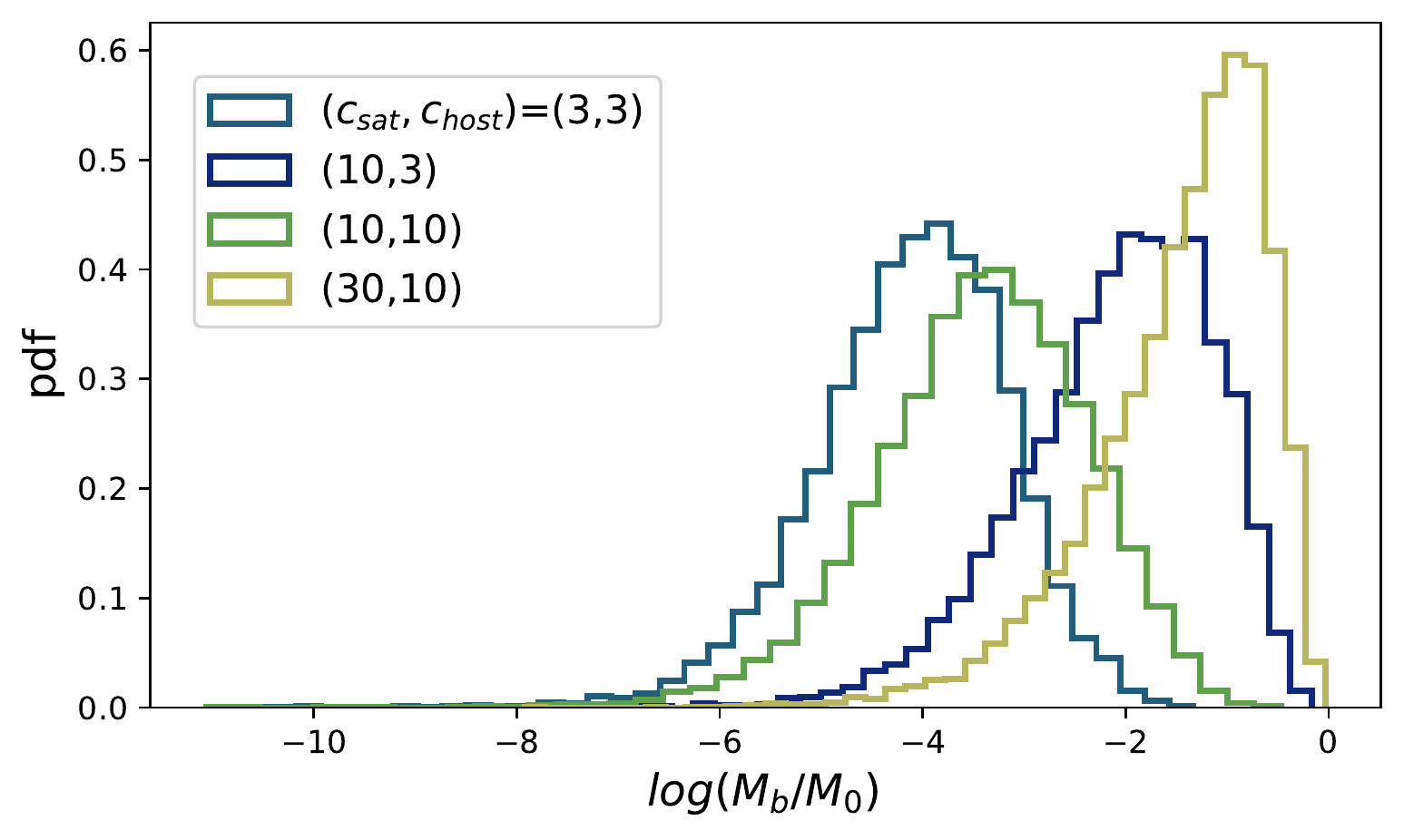}
\caption{The bound mass fraction predicted by our model for pairs of subhalo and host with 
concentrations $c_{\rm sat}$ and $c_{\rm host}$.
The distributions reflect the subhaloes' 
lognormal scatter around the displayed median concentration, and a distribution of pericentric
orbital distances as described in the text.}
\label{rpeqrh}
\end{figure}

\begin{figure*}
\centering
\includegraphics[width=.9\textwidth]{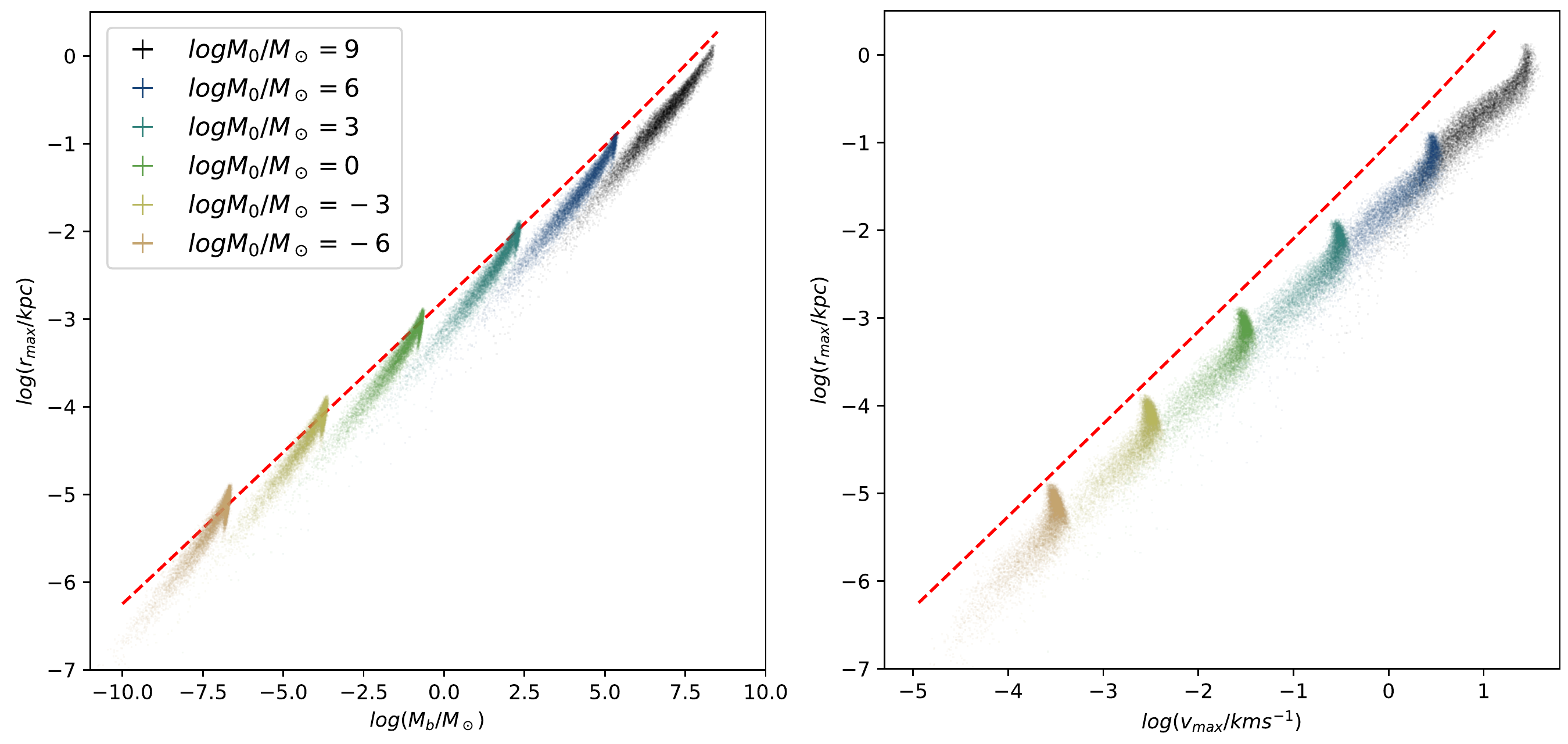}
\caption{The dependence of the characteristic radius $\rmx$
and velocity $\vmx$ with subhalo mass for a simplified model
of the accretion process (see text). The red dashed line in the left panel displays the characteristic radius $\rmx=2.16\rs$ for 
haloes on the same mass-concentration relation, at $z=0$. Similarly, the red-dashed line in the right panel shows the locus 
$(2.16\rs, 0.46\vs)$ for the same haloes. }
\label{rpeqrh}
\end{figure*}

\subsection{Dimensional scalings}
 
A full calculation of the abundance of CDM subhaloes based on our model requires making a number of assumptions on the population of haloes at accretion, as well as on the accretion process itself. We defer such a detailed analysis to future work. Here, we investigate how the characteristic scales $\rmx$ and $\vmx$
vary with the mass of the subhalo $\Mb$, based on a simplified model.

In order to do so, we require a prescription for the distribution of pericentric radii of accretion orbits in 
CDM. We adopt the distribution displayed in the Appendix of \citet{Wetzel11}. In terms of the virial radius $r_{\rm vir}$, 
this has a mean pericentric radius
of $\rp/r_{\rm vir}\approx0.32$ and a standard deviation of $0.23$. We assume that the same 
distribution is valid for all subhalo masses.
First, we explore what is the effect of this distribution on the 
final bound mass fractions. This is shown in Figure~10. As for Figure~9, we consider subhaloes with concentration $c_{\rm sat}$,
and hosts with concentration $c_{\rm host}$. The distributions 
shown in Fig.~10 reflect the distribution in pericentric radii,
together with the scatter in the mass concentration relation for the subhaloes (we use a lognormal scatter of $0.15~$dex; the properties of the host are kept fixed).
As mentioned earlier, higher subhalo concentrations allow it to
experience comparatively smaller levels of mass loss. All distributions, however span at least 3 orders of magnitudes in 
$\Mb/M_0$ between their 10\% and 90\% quantiles. Accretions at high redshifts, i.e. the case $c_{\rm sat}=c_{\rm host}=3$, display the most extreme levels of mass-loss.
The case of subhaloes of extremely low mass on MW-sized hosts 
is represented by the choice $(c_{\rm sat},c_{\rm host})=(30,10)$
and displays comparatively much higher mass fractions. 

We now fix a concentration of $c_{\rm}=8$ for the host, approximately representative of a MW-sized halo at around $z=1$. For the subhaloes, we assume the mass-concentration relation determined by \citet{Wang20},
 (we do not include the cutoff due to thermal effects). 
We use such concentrations to calculate the characteristic radii $\rs$ and densities $\rhs$ of the subhaloes at accretion, assuming an accretion redshift $z=1$, independent of mass. This provides us with all necessary ingredients to use our model to predict the long term properties of the same subhaloes: bound masses, $\Mb$,  characteristic radius $\rmx$ and velocity $\vmx$. 

The left panel of Figure~11 shows the distribution of tidally
stripped subhaloes in the plane of bound mass $\Mb$ and characteristic radius $\rmx$. We have considered subhaloes with masses at accretion $\log(M_0/M_\odot)\in\{9,6,3,0,-3,-6\}$. Each of these values 
correspond to a distribution of pairs $(\Mb, \rmx)$. As in Figure~10, this is the result
of the distribution of pericentric radii and of the  spread in the mass-concentration relation. Note that highly stripped subhaloes with identical initial mass are distributed along an $\rmx\sim\Mb^{1/2}$ locus, reflecting equation~(\ref{masssmall}). However, the global relation defined by the mix of different initial subhalo masses still follows the same dimensional scaling associated with isolated CDM haloes, $\rmx\sim\Mb^{1/3}$. This relation is shown by the red dashed line, which displays
the characterstic radius $\rmx=2.16\rs$ for haloes on the same mass concentration relation, at redshift $z=0$. Subhaloes appear to have somewhat lower characteristic radii, but a detailed quantification of this systematic shift would require a more 
accurate model for the mix of accreted haloes. 

The left panel of Fig.~11 shows the same tidally
stripped subhaloes in the plane of characteristic velocities $\vmx$ and characteristic radii $\rmx$. The red dashed line 
shows the location in this plane of haloes on the mass concentration
relation at $z=0$: $(\rmx,\vmx)=(2.16\rmx,0.46\vs)$. As for the 
relation between bound mass and characteristic radius, subhaloes appear to follow the same dimensional scaling of isolated CDM haloes: $\rmx\sim\vmx$. Subhaloes appear on average denser than isolated haloes, which is a well known effect \citep[e.g.,][]{Springel08}, but we defer the quantification of the shift between the two populations to 
future study.

\section{Discussion and Conclusions}

In this work, we have shown that the following two hypotheses
\begin{itemize}
    \item{a centrally divergent density profile $\rho\sim r^{-1}$;}
    \item{an isotropic phase space distribution}
\end{itemize}
are sufficient to ensure that CDM subhaloes orbiting CDM hosts can never be disrupted. This is valid 
for any mass of the subhalo at accretion. Therefore, this result implies that, like the halo mass function, the subhalo mass function of CDM models extends down to very low masses. For a 100 GeV WIMP model, very low masses means planet-sized masses \citep[][]{Green2005,Wang20}. Thanks to mass loss, 
subhalo mass functions extend even below that.

This follows from the realization that the tidal track 
of tidally evolving CDM subhaloes is very well described by the locus of
subhaloes that re-virialize after a truncation in energy space.
This is illustrated in Figure~2, and proves that tidal heating is essentially negligible for systems that satisfy the two conditions above: the bound regions of subhaloes evolved in 
previous numerical work have experienced negligible energy injection. In fact, this is the primary reason for the existence 
of a simple, uni-dimensional tidal track. The properties of cosmologically motivated haloes are such that the regime of very low mass subhaloes is not different: the relative importance of tidal heating 
does not increase for subhaloes with smaller masses at accretion -- see equation~(\ref{heatE0}). Therefore, subhaloes with any arbitrarily low initial mass are not more susceptible to heating, and evolve on the same tidal track.

This motivates building a simple model of tidal evolution which entirely neglects the effects of heating. In such a model, subhaloes are peeled in energy space and subsequently re-virialize. While a sharp truncation in energy may not be a fully accurate model of tidal stripping, it captures the fact that material is removed in an outside-in fashion. If material at high binding energies remains untouched while the outskirts are stripped, this model is bound to return the correct dimensional scalings of the structural properties of the evolved subhaloes.

While the tidal track is commonly
assumed to be parametrized by the bound mass fraction, within
this model it is parametrized by the value of the energy truncation $\nEt$, or equivalently, by the truncation radius $\rtr$. This better highlights the fact that, if a subhalo satisfies the two conditions above, it is possible to strip it down to arbitrarily small masses. A bound remnant will always ensue. This limit of small bound mass fractions is the limit $\nEt\to1$. We find that such truncations always result in bound remnants, proving that arbitrarily
small values of the bound mass fraction are indeed accessible, 
for any initial mass of the subhalo at accretion. 

In particular, the central power law index of the density profile, 
$\gamma=1$, fixes the slope of the tidal track in the regime of extreme mass loss: for low values of $\rmx/\rmx^0$ and $\vmx/\vmx^0$ we find $\vmx\sim\rmx^{1/2}$. 
Furthermore, satellites on such a tidal track 
grow progressively more resilient to tidal heating. This ensures that our model is consistent.

Using the classical, local description of the tidal field of the host, we connect the latter to the truncation energy.
This provides us with an estimate for the characteristic values $\rmx$ and $\vmx$ in the final phase of tidal evolution. These are the values along the tidal track that first allow the stripped subhalo to be almost entirely contained within its tidal radius. As the remnant approaches these structural properties, its evolution slows down substantially. 
We confirm this picture and test these estimates with a set of numerical experiments.  

A full quantification of the subhalo mass function based on our model is beyond the scope of this work. We have shown, however, that cosmological subhaloes experience significant degrees of mass loss, and that 
the spread in orbital properties and the scatter in the mass concentration relation result in wide distributions for the final bound mass fraction. In particular, subhaloes with higher satellite-to-host
mass ratios are due to lose a larger fraction of their
mass to stripping, as a consequence of their lower concentrations, even before dynamical friction is taken into account. 
Similarly, accretion events at higher redshifts, when the concentrations
of satellite and host are more similar to each other, 
result in lower bound mass fractions, with a median 
of $\Mb/M_0\approx10^{-4}$. In comparison, subhaloes 
of very low masses, and therefore high concentrations,
are significantly less affected, with a median of $\Mb/M_0\approx10^{-1}$ for the case $(c_{\rm sat}, c_{\rm host})=(30,10)$. 

While subhaloes with identical masses at accretion can lose very different fractions of their mass, the wide dynamic range covered by CDM subhaloes is such that the dimensional scaling between their characteristic properties remains tied to the scaling that describe isolated CDM haloes. More explicitly, $\rmx\sim \Mb^{1/3}$ and $\vmx\sim \rmx$. Subhaloes appear 
more dense, on average, than isolated haloes. However, a more detailed model of the accretion process would be required to quantify such a systematic shift in more detail. 

As mentioned in Section~3.2, it is possible that significantly under-concentrated subhaloes may be susceptible to tidal heating, and therefore escape from being described by our simple model. If this is the case, they may evolve following different trajectories in the $(\rmx/\rmx^0,\vmx/\vmx^0)$ plane, which will depend on the subhalo's orbit, effectively negating the intrinsic simplicity of a uni-dimensional tidal track. Similarly, the case of a host featuring a stellar disk should be considered separately. Our results, however, suggest that these factors may in fact not be crucial to the survival itself of the subhaloes, as this is directly connected to the survival of its density cusp. This, in turn, is effectively shielded from energy injection. The framework 
we have laid down will certainly facilitate further study on these outstanding questions.

Our results settle the longstanding issue of the properties and survival of CDM subhaloes. If the halo mass function of CDM haloes extends to masses of the order of $10^{-6}M_\odot$, so does the subhalo mass function.
In fact, our analysis directly shows why, despite the complexities of the tidal evolution process, the mass dependence of subhalo abundances is found to effectively track the one of halo abundances in cosmological simulations.
This results from the combination of the structural properties of CDM haloes (equation~(\ref{heatE1})) and the characteristic central divergence of their density profile. Even before further detailed analysis, our results imply that, as halo populations, CDM subhalo populations are indeed extremely abundant at low masses. 
In fact, it is not unlikely subhalo mass functions may be well approximated by simple extrapolation of current numerical results. 
This provides renewed confidence on the hope of using such small scales
properties to finally test the CMD model.

\section*{Software Citations}

This work used the following software packages:
\begin{itemize}

\item
Gadget2 \citep{Springel05}

\item 
\href{https://github.com/matplotlib/matplotlib}{\textt{matplotlib}} 
\citep{matplotlib}
\item 
\href{https://github.com/numpy/numpy}{\textt{NumPy}} 
\citep{numpy}

\item 
\href{https://www.python.org/}{\textt{Python}} 
\citep{python}
\item 
\href{https://github.com/scipy/scipy}{\textt{Scipy}}
\citep{scipy}
\end{itemize}

\section*{Data Availability}
The data underlying this article will be shared on reasonable request to the corresponding author.

\section*{Acknowledgements}
NCA is supported by an STFC/UKRI Ernest Rutherford Fellowship, Project Reference: ST/S004998/1.
I am always indebted to Adriano Agnello for countless stimulating conversations.

\appendix

\section{Functional parametrizations}

We report here a list of simple functional forms that describe our numerical results. 

The locus of the re-virialized energy truncated satellites displayed
in Fig.~2:
\begin{equation}
\log{\vmx\over\vmx^0}\approx {x\over 2}-{1\over4}\log(1+10^{2x})-{0.16\over{1+10^{2x+1.6}}}+0.075,
\end{equation}
where $x =\log({\rmx/\rmx^0}) $.

The quantities $\mu(\nEt)$, $\nu(\nEt)$, $\varpi(\nEt)$ displayed in Fig.~4:
\begin{gather}
    \log\mu(\nEt)\approx -{\nEt\over2}-0.225\nEt^5 \\
    \log\nu(\nEt)\approx -{\nEt\over2}-0.145\nEt^3+0.15\\
    \log\varpi(\nEt)\approx -0.15 - 0.2\log(1+10^{-5\nEt+2.9}).
\end{gather}

For the asymptotic relation between bound mass $\Mb$ and truncation radius $\rtr$, displayed in Fig.~5:
\begin{equation}
\log\Mb\approx 2\log\rtr - 1.1 .
\end{equation}

\bsp	
\label{lastpage}

\end{document}